\documentstyle[12pt]{article}

\input{psbox.tex}

\def\be{\begin{equation}}
\def\ee{\end{equation}}
\def\bea{\begin{eqnarray}}
\def\eea{\end{eqnarray}}

\def\br{\begin{thebibliography}{abc}}
\def\er{\end{thebibliography}}
\def\rf{\bibitem}
\def\cstars{C$^*$-algebras }
\def\cstar{C$^*$-algebra }
\def\sdp{\hbox{ \raisebox{.25ex}{\tiny $|$}\hspace{.2ex}{$\!\!\times $}} }

\def\a{\alpha}
\def\b{\beta}
\def\c{\raisebox{.4ex}{$\chi$}}
\def\d{\delta}
\def\e{\epsilon}
\def\f{\phi}
\def\g{\gamma}
\def\h{\eta}

\def\l{\lambda}
\def\m{\mu}

\def\p{\pi}
\def\q{\theta}
\def\r{\rho}
\def\s{\sigma}

\def\x{\xi}

\def\D{\Delta}

\def\ca{{\cal A}}
\def\cb{{\cal B}}
\def\cc{{\cal C}}

\def\ce{{\cal E}}

\def\ch{{\cal H}}

\def\ck{{\cal K}}

\def\co{{\cal O}}
\def\cp{{\cal P}}

\def\cu{{\cal U}}

\def\rt{\rightarrow}

\def\pa{\partial}
\def\del{\nabla}
\def\iff{\Leftrightarrow}

\def\bar#1{\overline{#1}}

\def\Hat#1{\rlap{\kern.10em$\widehat{\phantom G}$}#1}
\def\HAt#1{\rlap{\kern.05em$\widehat{\phantom G}$}#1}

\def\czp#1{\rlap{\kern.1em$\widehat{\phantom{G\vrule height.8em}}$}#1{}}
\def\Czp#1{\rlap{\kern.05em$\widehat{\phantom{G\vrule height.8em}}$}#1{}}

\newcommand{\sect}[1]{\setcounter{equation}{0}\section{#1}}
\newcommand{\subsect}[1]{\subsection{#1}}

\renewcommand{\thefootnote}{\fnsymbol{footnote}}
\def\fn{\footnote}
\def\sxn#1{\bigskip\medskip \sect{#1} \smallskip
                                                 }
\def\subsxn#1{\medskip \subsect{#1} \smallskip
                                                }

\begin{document}

\thispagestyle{empty}
\setcounter{page}{0}

\begin{flushright}
ICTP: IC/94/80\\
\ \hfill Napoli:  DSF-T-9/94\\
\ \hfill Syracuse: SU-4240-577\\
%\ \hfill May 1994
\end{flushright}

\vspace{.25cm}

\centerline {\LARGE NONCOMMUTATIVE LATTICES }
\vspace{5mm}
\centerline{\LARGE  AS FINITE APPROXIMATIONS }
\vspace{5mm}
%\centerline {\LARGE  AND THEIR NONCOMMUTATIVE GEOMETRIES}
\vspace{0.75cm}
\centerline {\large A.P. Balachandran$^1 $,
                    G. Bimonte$^{2,3}$,
                    E. Ercolessi$^1 $,
                    G. Landi$^{1,3,4,}$\fn[4]
{Fellow of the Italian National Council of Research (CNR) under Grant
No.~203.01.60. },}
\vspace{2mm}
\centerline{\large  F. Lizzi$^{3,5}$,
                    G. Sparano$^{3,5}$ and
                    P. Teotonio-Sobrinho$^1$}
\vspace{.50cm}
\centerline {\it $^1$ Department of Physics, Syracuse University,
Syracuse, NY 13244-1130, USA.}
\vspace{2mm}
\centerline {\it $^2$ International Centre for Theoretical Physics,
P.O. Box 586, I-34100, Trieste, Italy.}
\vspace{2mm}
\centerline {\it $^3$ INFN, Sezione di Napoli, Napoli, Italy.}
\vspace{2mm}
\centerline{\it $^4$ Dipartimento di Scienze Matematiche,
Universit\'a di Trieste,}
\centerline{\it P.le Europa 1, I-34127, Trieste, Italy.}
\vspace{2mm}
\centerline {\it $^5$ Dipartimento di Scienze Fisiche, Universit\`a di
Napoli,}
\centerline{\it Mostra d' Oltremare, Pad. 19, I-80125, Napoli, Italy.}

\vspace{.2cm}
\begin{abstract}
Lattice discretizations of continuous manifolds are common tools used in a
variety of physical contexts.
Conventional discrete approximations, however, cannot capture
all aspects of the original manifold, notably its topology.
In this paper we discuss an approximation scheme due to Sorkin
which correctly reproduces important topological aspects of
continuum physics.
The approximating topological spaces are partially ordered sets
(posets), the partial order encoding the topology.
Now, the topology of a manifold $M$ can be reconstructed from
the commutative C*-algebra $\cc(M)$ of continuous functions defined on it.
In turn, this algebra is generated by continuous probability densities
in ordinary quantum physics on $M$.
The latter also serve to
specify the domains of observables like the Hamiltonian.
For a poset, the role of this algebra is assumed by a
noncommutative C$^*$-algebra $\ca $.
This fact makes any poset a genuine `noncommutative' (`quantum')
space, in the sense that the algebra of
its `continuous functions' is a noncommutative C$^*$-algebra.
We therefore also have a remarkable connection between
finite approximations to quantum physics and noncommutative
geometries.
We use this connection to develop various approximation
methods for doing quantum physics using $\ca$.
\end{abstract}
\vfill {\sl To appear on The Journal of Geometry and Physics}
\newpage
\setcounter{page}{1}

\renewcommand{\thefootnote}{\arabic{footnote}}
\setcounter{footnote}{0}

\sxn{Introduction}\label{se:1}

Realistic physical theories
require approximations for the extraction of their predictions. A powerful
approximation method is the
discretization of continuum physics where manifolds are replaced by a
lattice of points. This discretization is
particularly effective for numerical work and
has acquired a central role in the study of
fundamental physical theories such as QCD \cite{DICK} or Einstein gravity
\cite{num}.

In these approximations, a manifold is typically substituted by a
 set of points with discrete topology. The latter is entirely incapable of
describing any significant topological attribute of the continuum, this
being equally the case for both local and global properties.
As a consequence, all
topological properties of continuum physical theories are lost.
For example, there is no nontrivial concept of winding number
on lattices with discrete topology
and hence also no way to associate solitons
with nonzero winding numbers in these approximations.

Some time ago, Sorkin \cite{So} studied
a very interesting method for
finite approximations of
manifolds by certain point sets in detail.
These sets are partially ordered sets
(posets) and have the ability to reproduce important topological
features of the continuum with remarkable fidelity.
[See also ref. \cite{Al}.]

Subsequent researches
\cite{BBET} developed these methods and made them usable for
approximate computations in quantum physics. They could thus become
viable alternatives to computational schemes like those in lattice QCD
\cite{DICK}.
This approximation scheme is briefly reviewed in
Section \ref{se:2}.

In this paper, we develop the poset approximation scheme
in a completely novel direction.

In quantum
physics on a manifold $M$, a fundamental role is played by the
\cstar $\cc(M)$ of continuous functions on $M$. Indeed, it is
possible to recover $M$, its topology and even its $C^\infty $-structure
when this algebra and a distinguished subalgebra are given
\cite{FD,Ma}. It is also possible to rewrite
quantum theories on $M$ by working exclusively with this algebra,
the tools for doing calculations efficiently also being readily
available \cite{Co,VG,CL}.
All this material on $\cc(M)$ is described in Section \ref{se:3}
with particular attention to its physical meaning.

In Section \ref{se:4} we show that the algebra $\ca $ replacing $\cc(M)$, when
$M$ is approximated by a poset, is an infinite-dimensional noncommutative
C$^*$-algebra. The poset and its topology are recoverable from the knowledge of
$\ca$. This striking result makes any poset a genuine `noncommutative'
(`quantum') space, in the sense that the algebra of its `continuous
functions' is a noncommutative C$^*$-algebra. This explains also our use of the
name `noncommutative lattices' for these objects\footnote{In the following we
will use the phrases `poset' and `noncommutative lattice' in an interchangeable
way.}.

We thus have a remarkable connection between topologically meaningful finite
approximations to quantum physics and noncommutative geometries. It bears
emphasis that this conclusion emerges in a natural manner while approximating
conventional quantum theory. Therefore the interest in noncommutative geometry
for a physicist need not depend on unusual space-time topologies like the one
used by Connes and Lott \cite{CL} in building the standard model. Furthermore,
these quantum models on posets are of independent interest and not just as
approximations to continuum theories, as they provide us with a whole class of
examples with novel geometries
\footnote{Dimakis and M\"{u}ller-Hoissen
\cite{DMH} have recently discussed a new approach to differential calculus and
noncommutative geometry on discrete sets which has interesting connections with
posets.}.

The \cstars for our posets are, as a rule, inductive limits of finite
dimensional matrix algebras, being examples of ``approximately finite
dimensional" algebras \cite{FD,Br,SV}. Therefore we can approximate $\ca$ by
finite dimensional algebras and in particular by a commutative finite
dimensional algebra $\cc(\ca)$. Their elements can be regarded as continuous
``functions" %(or rather, as sections of a certain bundle) on the poset. They
too encode the topology of the latter. The algebra $\cc(\ca)$ is also
strikingly simple, so that it is relatively easy to build a quantum theory
using $\cc(\ca)$. We describe these approximations in Section \ref{se:6}.

In Sections \ref{se:4.2} we discuss many aspects of quantum physics based on
$\ca$, drawing on known mathematical methods of the noncommutative geometer and
the C$^*$-algebraist.

Section \ref{se:7} deals with a concrete example having nontrivial topological
features, namely the poset approximation to a circle. We establish that global
topological effects can be captured by poset approximations and algebras
$\cc(\ca)$ by showing that the ``$\theta $-angle" for a particle on a circle
can also be treated using $\cc(\ca)$.

In Section \ref{se:5} we show how the \cstar for a poset can be generated by a
commutative subalgebra and a unitary group. We then argue that the algebra
$\cc(\ca)$ above can be recovered from this structural result and a gauge
principle.

The article concludes with some final remarks in Section \ref{se:9}.

\sxn{The Finite Topological Approximation}\label{se:2}

Let $M$ be a continuous topological space like, for example, the sphere $S^N$
or the Euclidean space ${\bf R}^N$. Experiments are never so accurate that they
can detect events associated with points of $M$, rather they only detect events
as occurring in certain sets $O_\l$. It is therefore natural to identify any
two points $x$, $y$ of $M$ if they can never be separated or distinguished by
the sets $O_\l$.

We assume that the sets
$O_\l$ cover $M$,
\be
   M=\bigcup _\l O_\l~, \label{2.1}
\ee
that each $O_\l$ is open and that
\be
   \cu =\{O_\l \} \label{2.3}
\ee
is a topology for $M$ \cite{EDM}.
This implies that both $O_\l\cup O_\m$ and
$O_\l\cap O_\m$ are in $\cu$ if $O_{\l,\m}\in \cu$.
This hypothesis is physically consistent because experiments can
isolate events in $O_\l\cup O_\m$ and $O_\l\cap O_\m$ if they can do so
in $O_\l$ and $O_\m$ separately, the former by detecting an event in either
$O_\l$ or $O_\m$, and the latter by detecting it in both $O_\l$ and $O_\m$.

Given $x$ and $y$ in $M$, we write $x\sim y$ if every set $O_\l$
containing either point $x$ or $y$ contains the other too:
\be
   x\sim y \mbox{  means  } x\in O_\l \iff y\in O_\l
{}~~~\mbox{for every}~~ O_\l~ . \label{2.2}
\ee
Then $\sim $ is an equivalence relation, and it is reasonable to replace $M$
by $M\, / \sim \equiv P(M)$ to reflect the coarseness of observations.
It is this
space, obtained by identifying equivalent points and equipped with the quotient
topology explained later, that will be our
approximation for $M$.

We assume that the number of sets $O_\l$ is finite when $M$ is compact so that
$P(M)$ is an approximation to $M$ by a finite set in this case. When $M$ is
not compact, we assume instead that each point has a neighbourhood
intersected by only finitely many $O_\l$ so that $P(M)$ is a ``finitary"
approximation to $M$ \cite{So}.
In the notation we employ, if $P(M)$ has $N$ points, we sometimes denote it
by $P_N(M)$.

The space $P(M)$ inherits the quotient topology from $M$ \cite{EDM}.This is
defined as follows.
Let $\Phi $ be the map from $M$ to $P(M)$ obtained by identifying equivalent
points. Then a set in $P(M)$ is declared to be open if its inverse image for
$\Phi$ is open in $M$. The topology generated by these open sets is the finest
one compatible with the continuity of $\Phi$.

Let us illustrate these considerations for a cover of $M=S^1$ by four open
sets as in Fig. 1(a).
In that figure, $O_{1,3}\subset O_2\cap O_4$.
Fig. 1(b) shows the corresponding discrete space $P_4(S^1)$, the
points $x_i$ being images of sets in $S^1$. The map $\Phi\, :\, S^1\rightarrow
P_4(S^1)$ is given by
$$
  O_1 \rightarrow x_1, ~ ~ ~ ~ ~
        O_2\setminus [O_2\cap O_4] \rightarrow x_2~,
$$
\be
  O_3 \rightarrow x_3, ~ ~ ~ ~ ~
        O_4\setminus [O_2\cap O_4] \rightarrow x_4 ~ .\label{2.3a}
\ee
The quotient topology for $P_4(S^1)$ can be read off from Fig. 1,
the open sets being
\be
\{x_1\}~,  ~ \{x_3\}~,  ~\{x_1,x_2,x_3\}~,  ~ \{x_1,x_4,x_3\}~,
 \label{2.4}
\ee
and their unions and intersections (an arbitrary number of the latter being
allowed as $P_4(S^1)$ is finite).

\begin{figure}[hbtp]
\begin{center}
\mbox{\psboxto(0cm;7cm){fig1.eps}}
\end{center}
{\footnotesize{\bf Fig.1.} (a) shows an open cover for the circle $S^1$
and (b) the resultant discrete space $P_4(S^1)$. $\Phi $ is the map
(\ref{2.3a}). }
\end{figure}

Notice that our assumptions
allow us to isolate events in certain sets of the form
$O_\l\setminus [O_\l\cap O_\m]$ which may not be open. This means that there
are in general points in $P(M)$ coming from sets which are not open in $M$ and
therefore are not open in the quotient topology.

Now in a Hausdorff space \cite{EDM},
for any two distinct points $x$ and $y$ there
exist open sets $O_x$ and $O_y$, containing $x$ and $y$ respectively,
such that $O_x\cap O_y=\emptyset $. A finite Hausdorff space necessarily has
the
discrete topology and hence each of its points is an open set.
So $P(M)$ is not Hausdorff. However, it can be shown \cite{So} that it is a
$T_0$ space \cite{EDM}. $T_0$ spaces are defined as spaces in
which, for any two distinct points, there is an
open set containing at least one of these points and not the other.
For example, given the points
$x_1$ and $x_2$ of $P_4(S^1)$, the open set $\{x_1 \}$ contains $x_1$ and
not $x_2$, but there is no open set containing $x_2$ and not $x_1$.

In $P(M)$, we can introduce a partial order $\preceq $ \cite{Al,HW,St}
by declaring that:
\[ x\preceq y \mbox{ if every open set containing } y
\mbox{ contains also } x~.\]
$P(M)$ then becomes a \underline{partially ordered set}
or a \underline{poset}.
Later, we will write $x\prec y$ to indicate that $x\preceq y$ and $x\neq y$.

Any poset can be represented by a Hasse diagram constructed by arranging its
points at different levels and connecting them using the
following rules:
\begin{enumerate}
\item[1)] if $x\prec y$, then $x$ is at a lower level than $y$;
\item[2)] if $x\prec y$ and there is no $z$ such that $x\prec z\prec y$, then
$x$ is
at the level immediately below $y$ and these two points
are connected by a line called a link.
\end{enumerate}

For $P_4(S^1)$, the partial order reads
\be
x_1\preceq x_2~,~~ x_1\preceq x_4~,~~ x_3\preceq x_2~,~~
x_3\preceq x_4~,~~ \label{2.5}
\ee
where we have omitted writing the relations $x_j\preceq x_j$.
The corresponding Hasse diagram is shown in Fig. 2.

\begin{figure}[hbtp]
\begin{center}
\mbox{\psboxto(0cm;5cm){fig2.eps}}
\end{center}
\begin{center}
{\footnotesize{\bf Fig.2.} The Hasse diagram for the circle poset
$P_4(S^1)$. }
\end{center}
\end{figure}

In the language of partially ordered sets, the smallest open set $O_x$
containing a point $x \in P(M)$ consists of all $y$ preceding
$x$: $O_x = \{y \in P(M) \; : \; y \preceq x \}$. In the Hasse diagram, it
consists of $x$ and all points we encounter as we travel along links
from $x$ to the bottom. In Fig. 2, this rule gives
$\{x_1,x_2,x_3\}$ as the smallest open set containing $x_2$, just as in
(\ref{2.4}).

As another example, Fig. 3 shows a cover of $S^1$ by $2N$ open sets $O_j$ and
the Hasse diagram of its poset $P_{2N}(S^1)$.

\begin{figure}[hbtp]
\begin{center}
\mbox{\psboxto(0cm;7cm){fig4.eps}}
\end{center}
{\footnotesize{\bf Fig.3.} In (a) is shown a covering of $S^1$ by open sets
$O_j$ with $O_3=O_2\cap O_4$, $O_5=O_4\cap O_6$, ...,$O_1=O_{2N}\cap O_2$.
(b) is the Hasse diagram of its poset. }
\end{figure}

As one example of a three-level poset, consider the Hasse diagram of Fig. 4
for a finite approximation $P_6(S^2)$ of the two-dimensional sphere $S^2$
derived in \cite{So}. Its open sets are generated by
$$
\{x_1\}~,  ~ ~ \{x_3\}~,  ~ ~ \{x_1,x_2,x_3\}~,
{}~ ~ \{x_1,x_4,x_3\}~,
$$
\be
\{x_1,x_2,x_5,x_4,x_3\}~, ~ ~ \{x_1,x_2,x_6,x_4,x_3\}~,
{}~ ~ \label{2.6}
\ee
by taking unions and intersections.

\begin{figure}[hbtp]
\begin{center}
\mbox{\psboxto(0cm;5cm){fig3.eps}}
\end{center}
\begin{center}
{\footnotesize{\bf Fig.4.} The Hasse diagram for the two-sphere poset
$P_6(S^2)$.}
\end{center}
\end{figure}

We conclude this section by recalling that one of the most remarkable
properties of a poset is its ability to accurately reproduce the fundamental
group \cite{fun} of the manifold it approximates. For example, as for $S^1$,
the fundamental group of $P_N(S^1)$ is ${\bf Z}$ whenever $N \geq 4$ \cite{So}.
It is this property that allowed us to argue in \cite{BBET} that global
topological information relevant for quantum physics can be captured by such
discrete approximations. We will show this result again in Section \ref{se:7}.
There we will consider the ``$\theta$-angle" quantisations of a particle moving
on $S^1$ and establish that they can also be recovered when $S^1$ is
approximated by $P_{N}(S^1 )$. This result will be demonstrated by constructing
suitable line bundles on $P_{N}(S^1)$.

\sxn{Topology from Quantum Physics}\label{se:3}

In conventional quantum physics, the configuration space is generally a
manifold when the number of degrees of freedom is finite. If $M$ is
this
manifold and $\ch$ the Hilbert space of wave functions, then $\ch$
consists of all square integrable functions on $M$ for a suitable
integration measure. A wave function $\psi $ is only required to be
square integrable. There is no need for $\psi $ or the probability
density $\psi ^*\psi $ to be a continuous function on $M$. Indeed there
are plenty of noncontinuous $\psi $ and $\psi ^*\psi $. Wave functions
of course are not directly observable, but probability densities are,
and the existence of noncontinuous probability densities have
potentially disturbing implications. If all states of the system are
equally available to preparation, which is the case if all self-adjoint
operators are equally observable, then clearly we cannot infer the
topology of $M$ by measurements of probability densities.

It may also be recalled in this connection that any two infinite-dimensional
(separable) Hilbert spaces $\ch_1$ and $\ch_2$ are unitarily related. [Choose
an orthonormal basis $\{h_n^{(i)}\}$, $(n=0,1,2,...)$ for $\ch_i$ $(i=1,2)$.
Then a unitary map $U:\ch_1\rightarrow \ch_2$ from $\ch_1$ to $\ch_2$ is
defined by $Uh_n^{(1)}=h_n^{(2)}$.] They can therefore be identified, or
thought of as the same. Hence the Hilbert space of states in itself contains no
information whatsoever about the configuration space.

It seems however that not all self-adjoint operators have equal status in
quantum theory. Instead, there seems to exist a certain class of privileged
observables $\cp \co$ which carry information on the topology of $M$ and also
have a special role in quantum physics. This set $\cp \co$ contains operators
like the Hamiltonian and angular momentum and particularly also the set of
continuous functions $\cc(M)$ on $M$, vanishing at infinity if $M$ is
noncompact.

In what way is the information on the topology of $M$ encoded in $\cp \co$? To
understand this, recall that an unbounded operator such as a typical
Hamiltonian $H$ cannot be applied on all vectors in $\ch$. Instead, it can be
applied only on vectors in its domain $D(H)$, the latter being dense in $\ch$
\cite{hil}. In ordinary quantum mechanics, $D(H)$ typically consists of
twice-differentiable functions on $M$ with suitable fall-off properties at
$\infty $ in case $M$ is noncompact. In any event, what is important to note is
that if $\psi $, $\chi \in D(H)$ in elementary quantum theory, then $\psi
^*\chi \in \cc(M)$. A similar property holds for the domain $D$ of any
unbounded operator in $\cp \co$: if $\psi $, $\chi \in D$, then $\psi ^*\chi
\in \cc(M)$. It is thus in the nature of these domains that we must seek the
topology of $M$\footnote{Our point of view about the manner in which topology
is inferred from quantum physics was developed in collaboration with G. Marmo
and A. Simoni.}.

We have yet to remark on the special physical status of $\cp \co$ in quantum
theory. Let $\ce$ be the intersection of the domains of all operators in $\cp
\co$. Then it seems that the basic physical properties of the system, and even
the nature of $M$, are all inferred from observations of the privileged
observables on states associated with $\ce$\footnote{Note in this connection
that any observable of $\cp \co$ restricted to $\ce$ must be essentially
self-adjoint \cite{hil}. This is because if significant observations are all
confined to states given by $\ce$, they must be sufficiently numerous to
determine the operators of $\cp \co$ uniquely.}.

This discussion shows that for a quantum theorist, it is quite important to
understand clearly how $M$ and its topology can be reconstructed from the
algebra $\cc(M)$. Such a reconstruction theorem already exists in the
mathematical literature. It is due to Gel'fand and Naimark \cite{FD}, and is a
basic result in the theory of C$^*$-algebras and their representations. Its
existence is reassuring and indicates that we are on the right track in
imagining that it is $\cp \co$ which contains information on $M$ and its
topology.

We should point out the following in this regard however: it is not clear
that the specific mathematical steps one takes to reconstruct the
manifold from the algebra have a counterpart in the physical
operations done to reconstruct it from observations.

Let us start by recalling that a
\cstar $\ca$, commutative or otherwise, is an algebra with a norm
$\parallel \cdot \parallel$ and an antilinear involution * such that
$\parallel a\parallel =\parallel a^*\parallel $, $\parallel
a^*a\parallel =$\mbox{$\parallel a^*\parallel ~ \parallel a\parallel $} and
$(ab)^*=b^*a^*$ for $a,b\in \ca $. The algebra $\ca$ is also assumed to be
complete in the given norm.

Examples of \cstars  are:
\begin{itemize}
\item[1)] The (noncommutative) algebra of $n\times n$ matrices $T$ with
$T^*$ given by the
hermitian conjugate of $T$ and the squared norm $\parallel T\parallel^2$
being equal to the largest eigenvalue of $T^* T$;
\item[2)] The (commutative) algebra $\cc(M)$ of continuous functions
on a Hausdorff topological space $M$
(vanishing at infinity if $M$ is not compact),
with * denoting complex conjugation and the norm given by
the supremum norm, $\parallel f \parallel =\sup_{x\in M}|f(x)|$.
\end{itemize}

It is the latter example,
establishing that we can associate a commutative \cstar to a Hausdorff
space, which is relevant for the Gel'fand-Naimark theorem.
The Gel'fand-Naimark results then show how, given
{\it any} commutative \cstar $\cc$, we can reconstruct a Hausdorff
topological space $M$ of which $\cc$ is the algebra of continuous functions.

We now explain this theorem briefly. Given such a $\cc$, we let $M$ denote the
space of equivalence classes of irreducible representations
(IRR's)\footnote{The trivial IRR  given by $\cc\rightarrow \{0\}$ is not
included in $M$. It will therefore be ignored here and hereafter.}, also called
the structure space, of $\cc$\footnote{Some readers might be more familiar with
a slightly different construction, where $M$ is taken to be the space of
maximal two-sided ideals of $\cc$ instead of the space of irreducible
representations. These two constructions agree because for a commutative
\cstar, not only are the kernels of irreducible representations maximal
two-sided ideals, but also any maximal two-sided ideal is the kernel of an
irreducible representation \cite{FD}.}. The \cstar $\cc$ being commutative,
every IRR is one-dimensional. Hence, if $x\in M$ and $f \in \cc$, the image
$x(f)$ of $f$ in the IRR defined by $x$ is a complex number. Writing $x(f)$ as
$f(x)$, we can therefore regard $f$ as a complex-valued function on $M$ with
the value $f(x)$ at $x\in M$. We thus get the interpretation of elements in
$\cc$ as {\bf C}-valued functions on $M$.

We next topologise $M$ by declaring a subset of $M$ to be closed
if it is the set of zeros of some  $f\in \cc$. (This is natural to do
since the set of zeros of a continuous function is closed.)
The topology of $M$ is generated by these closed sets, by taking
intersections and finite unions. It is called the hull kernel or
Jacobson topology \cite{FD}.

Gel'fand and Naimark then show that the algebra $\cc(M)$
of continuous functions on $M$ is isomorphic to the starting algebra $\cc$.
It is therefore the case that the commutative
\cstar $\cc$ which reconstructs a given $M$ in the above fashion is unique.
Also the requirement $\cc=\cc(N)$ uniquely fixes $N$ up
to homeomorphisms. In this way, we recover a
topological space $M$, uniquely up to homeomorphisms, from the algebra
$\cc$\footnote{ We remark that more refined attributes of $M$ such as a
C$^\infty $-structure, can also be recovered using only algebras if more
data are given. For the C$^\infty $-structure, for example, we must also
specify an appropriate subalgebra $\cc ^\infty(M) $ of $\cc(M)$.
The C$^\infty
$-structure on $M$ is then the unique C$^\infty $-structure for which
the elements of $\cc^\infty(M)$ are all the C$^\infty $-functions
\cite{Ma}.}.

We next briefly indicate how we can do quantum theory starting from
$\cc(M)=\cc$.

Elements of $\cc$ are observables, they are not quite wave functions.
The set of all wave functions forms a Hilbert space $\ch$. Our first step
in constructing $\ch$, essential for quantum physics, is the
construction of the space $\ce$ which will serve as the common domain of
all the privileged observables.

The simplest choice for $\ce$ is $\cc$ itself
\footnote{Differentiability
requirements will in general further restrict $\ce$. As a rule we will ignore
such details in this article.}.
With this choice, $\cc$ acts on $\ce$, as $\cc$
acts on itself by multiplication. The presence of this action is important as
the privileged observables must act on $\ce$. Further, for $\psi, ~ \chi \in
\ce$, $\psi ^*\chi \in \cc$, exactly as we want.

Now Gel'fand and Naimark have established that it is possible to
define an integration measure $d\mu$ over the structure space $M$ of $\cc$,
such that every $f \in \cc$ has a finite integral. A scalar product
$(\cdot ,\cdot )$ for elements of $\ce$ can therefore be defined by
setting
\be
   (\psi ,\chi )=\int _M d\mu (x) (\psi ^*\chi )(x)\;.\label{3.1}
\ee
The completion of the space $\ce$ using this scalar product gives the
Hilbert space $\ch$.

The final set-up for quantum theory here is conventional. What is novel
is the shift in emphasis to the algebra $\cc$. It is from this algebra
that we now regard the configuration $M$ and its topology as having been
constructed.

There is of course no reason why $\ce$ should always be $\cc$. Instead
it can consist of sections of a vector bundle over $M$ with a
$\cc$-valued positive definite sesquilinear form $<\cdot ,\cdot >$.
(The form $<\cdot ,\cdot >$ is positive definite if $<\a,\a>$ is a
nonnegative function for
any $\a\in \ce$, which identically vanishes iff $\a=0$.) The scalar
product is then written as
\be
        (\psi ,\chi )=\int_Md\mu (x)<\psi ,\chi >(x).\label{3.2}
\ee
The completion of $\ce$ using this scalar product as before gives $\ch$.

\sxn{The Noncommutative Geometry of a Noncommutative Lattice}\label{se:4}

\subsxn{The Noncommutative Algebra of a Noncommutative Lattice}\label{se:4.1}

In the preceding sections we have seen how a commutative C$^*$-algebra
reconstructs a Hausdorff topological space. We have also seen that a poset
is not Hausdorff. It cannot therefore be reconstructed from a commutative
\cstar. It is however possible to reconstruct it,
and its topology, from a {\em noncommutative} \cstar.

Let us first recall a few definitions and results from operator theory
\cite{hil} before outlining this reconstruction theorem. An operator in a
Hilbert space is said to be of finite rank if the orthogonal complement
of its null space is finite dimensional. It is thus essentially like a
finite dimensional matrix as regards its properties even if the Hilbert
space is infinite dimensional. An operator $k$ in a Hilbert space is
said to be compact if it can be approximated arbitrarily closely in norm
by finite rank operators. Let $\l_1,\l_2,$ ... be the eigenvalues of
$k^* k$ for such a $k$, with $\l_{i+1} \leq \l_i$ and
an eigenvalue of multiplicity $n$
occurring $n$ times in this sequence. (Here and in what follows, $*$
denotes the adjoint for an operator.)
Then $\l_n\rightarrow 0$ as
$n\rightarrow \infty $. It follows that the operator $1\!\!1$ in an infinite
dimensional Hilbert space is not compact.

The set $\ck$ of all compact operators $k$ in a Hilbert space is a \cstar.
It is a two-sided ideal in the \cstar $\cb$ of all bounded operators
\cite{FD,Si}.

Note that the sets of finite rank, compact and bounded operators are all
the same in a finite dimensional Hilbert space. All operators in fact
belong to any of these sets in finite dimensions.

The construction of $\ca$ for a poset rests on the following result from the
representation theory of $\ck$. The representation of $\ck$ by itself is
irreducible \cite{FD} and it is the \underline{only} IRR of $\ck$ up to
equivalence.

The simplest nontrivial poset is $P_2=\{p,q\}$ with $q\prec p$. It is
shown in Fig. 5. It is the poset for the interval $[r,s]$ ($r<s$) where the
latter is covered by the open sets $[r,s[$ and $[r,s]$.
The map from subsets of $[r,s]$ to the points of $P_2$ is
\be
\{s\} \rightarrow p~, ~~~[r,s[ \rightarrow q~.\label{4.1.1}~
\ee

The algebra $\ca$ for $P_2$ is
\be
        \ca= {\bf C}1\!\!1 + \ck =
             \{ \l 1\!\!1 + k \, : \, \l\in {\bf C}, k \in \ck \}
\;,\label{4.1}
\ee
the Hilbert space on which
the operators of $\ca$ act being infinite dimensional.

We can see this result from the fact that
$\ca$ has only two IRR's and they are
given by
\bea
        p:\l 1\!\!1 +k& \rightarrow & \l ~,
\nonumber \\
q:\l 1\!\!1 +k& \rightarrow & \l 1\!\!1 +k ~.
\label{4.1a}
\eea
This remark about IRR's becomes plausible if it is
remembered that $\ck$ has only one IRR.

Thus the structure space of $\ca$ has only two points $p$ and $q$.
An arbitrary element $\l 1\!\!1 +k$ of $\ca$ can be regarded as a ``function"
on it if, in analogy to the commutative case, we set
\bea
(\l 1\!\!1 +k) (p) &:=& \l \; \nonumber \\
(\l 1\!\!1 +k) (q) &:=& \l 1\!\!1 +k \; . \label{4.1b}
\eea
Notice that in this case the function $\l 1\!\!1 +k$ is not valued
in {\bf C} at all points. Indeed, at different points it is valued
in different spaces, {\bf C} at $p$ and a subset of bounded operators
on an infinite Hilbert space at $q$\footnote{Such an interpretation of $\ca$
as functions on the poset can also be stated in a more rigorous way.
In a paper under preparation, we will in fact show that $\ca$ is isomorphic
to the algebra of continuous sections of a suitable bundle over the poset, in
the same way that the algebra of continuous functions on a manifold $M$
is isomorphic to the algebra of continuous
sections of the trivial one-dimensional complex vector bundle on $M$.}.

\begin{figure}[hbtp]
\begin{center}
\large
\mbox{\psannotate{\psboxto(0cm;5cm){fig6.eps}}{ %\fillinggrid
\at(5\pscm;4\pscm){$q$}
\at(4.5\pscm;11.5\pscm){$p$}
\at(19\pscm;4\pscm){$(\l 1\!\!1 +k)(q)=\l 1\!\!1 +k$}
\at(19\pscm;11\pscm){$(\l 1\!\!1 +k)(p)=\l$}}}
\normalsize
\end{center}
{\footnotesize{\bf Fig.5.} (a) is the poset for the interval $[r,s]$
when covered by the open sets $[r,s[$ and $[r,s]$. (b) shows the values of
a generic element $\l 1\!\!1 +k$ of its algebra $\ca$  at its two points
$p$ and  $q$.}
\end{figure}

Now we can use the hull kernel topology for the set $\{p,q\}$.
For this purpose, consider the function $k$. It vanishes at $p$ and
not at $q$, so $p$ is closed. Its complement $q$ is hence open. So of
course is the whole space $\{p,q\}$. The topology of $\{p,q\}$ is thus given
by Fig. 5(a) and is that of the $P_2$ poset just as we want.

We remark here that for finite structure spaces one can equivalently define
the Jacobson topology as follows. Let $I_x$ be the kernel for the
IRR $x$. It is the (two-sided) ideal mapped to 0 by the IRR $x$. We set
$x\prec y$ if $I_x\subset I_y$ thereby converting the space of IRR's
into a poset. The topology in question is the topology of this poset.
In our case, $I_p=\ck $, $I_q=\{0\}\subset I_p$ and hence $q\prec p$. This
gives again Fig. 5(a).

Hereafter in this paper, by `ideals' we always mean two--sided ideals.

We next consider the $\bigvee$ poset. It can be obtained from the
following open cover of the interval $[0,1]$:
\bea
& & [0,1] = \bigcup_\l \co_\l~, \nonumber \\
& & \co_1 = [0, 2/3[~,~~ \co_2 = ]1/3, 1]~,~~ \co_3 = ]1/3, 2/3[~.
\label{4.2.1}
\eea
The map from subsets of $[0,1]$ to the points of the $\bigvee$ poset
in Fig. 6(a) is given by
\be
[0, 1/3] \rightarrow \a~,~~ ]1/3, 2/3[ \rightarrow \g~,~~
[2/3, 1] \rightarrow \b~.
\label{4.2.2}
\ee

Let us now find the algebra $\ca $ for the $\bigvee$ poset. This poset
has two arms 1 and 2. The first step in the construction is to attach an
infinite-dimensional Hilbert space $\ch_i$ to each arm $i$ as
shown in Fig. 6(a).
Let $\cp_i$ be the orthogonal projector on $\ch_i$ in $\ch_1\oplus
\ch_2$ and $\ck_{12}=\{k_{12}\}$ be the set of all compact operators in
$\ch_1\oplus \ch_2$. Then \cite{FD}
\be
        \ca= {\bf C}\cp_1 + {\bf C}\cp_2 + \ck_{12} ~ .\label{4.2}
\ee

\begin{figure}[hbtp]
\begin{center}
\large
\mbox{\psannotate{\psboxto(0cm;6cm){fig7.eps}}{ %\fillinggrid
\at(7\pscm;4\pscm){$\g $}
\at(4\pscm;7\pscm){$\ch_1 $}
\at(8.5\pscm;7\pscm){$\ch_2 $}
\at(3.6\pscm;11\pscm){$\a $}
\at(8\pscm;11\pscm){$\b $}
\at(14.2\pscm;4\pscm){$a(\g)=\l_1\cp_1+\l_2\cp_2+k_{12} $}
\at(14.7\pscm;11\pscm){$a(\a)=\l_1 $}
\at(19.5\pscm;11\pscm){$a(\b)=\l_2 $}
}}
\normalsize
\end{center}
{\footnotesize{\bf Fig.6.} (a) shows the $\bigvee$ poset and the
association of an infinite
dimensional Hilbert space $\ch_i$ to each of its arms.
(b) shows the values of a typical element $a=\l_1\cp_1+\l_2\cp_2+
k_{12}$ of its algebra at its three points. }
\end{figure}

The IRR's of $\ca$ defined by the three points of the poset are given by
Fig. 6(b). It is easily seen that the hull kernel topology correctly
gives the topology of the $\bigvee$ poset.

The generalization of this construction to any (connected) two-level poset
is as follows. Such a poset is composed of several $\bigvee$'s.
Number the arms and attach an
infinite dimensional Hilbert space $\ch_i$ to each arm $i$
as in Figs. 7(a) and 7(b).
To a $\bigvee$ with arms
$i,i+1$, attach the algebra $\ca_i$ with elements
$\l_i\cp_i+\l_{i+1}\cp_{i+1}+k_{i,i+1}$. Here $\l_{i},\l_{i+1}$ are any two
complex numbers, $\cp_i ,\cp_{i+1}$
are orthogonal projectors on $\ch_i$, $\ch_{i+1}$ in the Hilbert
space $\ch_i\oplus \ch_{i+1}$ and $k_{i,i+1}$ is any compact operator
in $\ch_i\oplus \ch_{i+1}$. This is as before. But now, for glueing
the various $\bigvee$'s together, we also impose the
condition $\l_j=\l_k$ if the lines $j$ and $k$ meet at a top point. The algebra
$\ca$ is then the direct sum of $\ca_i$'s with this condition:
\be
        \ca= \bigoplus \ca_i ~ ~
\ca_i = \l_i\cp_i+\l_{i+1}\cp_{i+1}+k_{i,i+1} \label{4.3}
\ee
\[ \mbox{with } \l_j=\l_k \mbox{ if lines } j,k
\mbox{ meet at top.}    \]

Figs. 7(a) and (b) also show the values of an element $a=\oplus
[\l_i\cp_i+\l_{i+1}\cp_{i+1}+k_{i,i+1}]$ at the different points of two
typical two-level posets.

\begin{figure}[hbtp]
\begin{center}
\mbox{\psboxto(0cm;7cm){fig8.eps}}
\end{center}
{\footnotesize{\bf Fig.7.} These figures show how the Hilbert spaces $\ch_i$
are attached to the arms of two two-level posets. They also show
the values of a generic member
$a$ of their algebras $\ca$ at their points.}
\end{figure}

There is a systematic construction of $\ca$ for any poset (that is, any
``finite $T_0$ topological space") which generalizes the preceding
constructions for two-level posets. It is explained in the book by
Fell and Doran
\cite{FD} and will not be described here.

It should be remarked that actually
the poset does not uniquely fix its algebra as there are in general many
non-isomorphic (noncommutative) C$^*$-algebras with the same poset
as structure space \cite{BL}. This is to be contrasted with
the Gel'fand-Naimark result asserting that the (commutative)
\cstar associated to a
Hausdorff topological space (such as a manifold) is unique.
 The Fell-Doran choice of the algebra for the poset seems to be
the simplest. We will call it $\ca$ and adopt it in this paper.

\subsxn{Finite Dimensional and Commutative Approximations}\label{se:6}

In general, the algebra $\ca$ is infinite dimensional.
This makes it difficult to use  it in explicit calculations, notably
in numerical work.

We will show that there is a natural
sequence of finite dimensional approximations to the algebra $\ca$
associated to a poset. For two-level posets,
the leading nontrivial approximation here is
commutative while the succeeding ones are not.
In this case, the commutative approximation
$\cc(\ca)$ has a suggestive physical interpretation.
Further these approximations correctly capture the topology of the
poset and can thus provide us with excellent models to
initiate practical calculations, and to gain experience and insight
into noncommutative geometry in the quantum domain.

The existence of these finite dimensional approximations relies on a
remarkable property that characterizes the \cstar $\ca$ associated to
a poset, namely the fact that $\ca$ is an approximately finite
dimensional (AF) algebra \cite{Br}.
Technically this means that $\ca$ is an inductive limit \cite{FD} of
finite dimensional \cstars (that is, direct sums of matrix algebras).

Incidentally, we remark here that there exists a construction
to obtain such sequence of finite dimensional algebras directly
from the topology of the poset. It is explained in \cite{Br} and is
based on the possibility of associating a diagram, the so-called
Bratteli diagram, to any finite $T_0$ space. This construction also gives
a new way, different from
the method of Fell and Doran discussed in the previous
section, to obtain the algebra $\ca$ of a poset.
We will not describe it here. Instead,
we limit ourselves to discussing only a few examples.

Let us start with the two-point poset of Fig. 5(a). The algebra
associated to it is $\ca= {\bf C}1\!\!1+ \ck$.
Consider the following sequence of \cstars of increasing
dimensions, the $*$-operation being hermitian conjugation:
\bea
& & \ca_0 = {\bf C}~, \nonumber \\
& & \ca_1 = M(1, {\bf C}) \oplus {\bf C}~, \nonumber \\
& & \ca_2 = M(2, {\bf C}) \oplus {\bf C}~, \nonumber \\
& & \cdots \nonumber \\
& & \ca_n = M(n, {\bf C}) \oplus {\bf C}~, \nonumber
\\ & & \ldots~~~~~~~~~~~~~
\label{6.1}
\eea
where $M(n, {\bf C})$ is the the C$^*$-algebra
of $n \times n$ complex matrices. A typical element of $\ca_{n}$ is
\be
a_{n} = \left[
\begin{array}{cc}
m_{n \times n} &0 \\
0 &\l
\end{array}
\right]~,
\label{6.2}
\ee
where $m_{n \times n}$ is an $n \times n$ complex matrix and
$\l$ is a complex number. Note that the subalgebra $M(n,{\bf C})$
consists of matrices of the form (\ref{6.2}) with the last row
and column zero.

The algebra $\ca_n$ is seen to approach $\ca$ as $n$ becomes larger
and larger. We can make this intuitive observation more precise.
There is an inclusion
\be
F_{n+1,n} ~:~ \ca_n \rt \ca_{n+1}
\label{6.3}
\ee
given by
\be
a_{n} = \left[
\begin{array}{cc}
m_{n \times n} & 0 \\
0 &\l
\end{array}
\right]~~ \rt ~~
\left[
\begin{array}{ccc}
m_{n \times n} & 0 & 0 \\
0 & \l & 0 \\
0 & 0 & \l
\end{array}
\right]~.
\label{6.4V}
\ee
It is a $*$-homomorphism \cite{FD} since
\be
F_{n+1,n}(a^{*}_{n}) =
[F_{n+1,n}(a_{n})]^{*}~.
\label{6.5}
\ee
Thus the sequence
\be
\ca_0 \rt \ca_1 \rt \ca_2 \rt \cdots
\label{6.6}
\ee
gives a directed system of C$^*$-algebras. Its inductive limit
is $\ca$ as is readily proved using the definitions in
\cite{FD}.

We must now associate appropriate representations to $\ca_n$ which
will be good approximations to the two-point poset.

The algebra $\ca_1$ is trivial. Let us ignore it. All the remaining
algebras $\ca_n$ have the following two representations:
\begin{enumerate}
\item[a)] The one-dimensional representation $p_n$ with
\be
p_n : a_n \rt \l~.
\label{6.7}
\ee
\item[b)] The defining representation $q_n$ with
\be
q_n : a_n \rt a_n~.
\label{6.8}
\ee
\end{enumerate}

It is clear that these representations approach the representations $p$
and $q$ of $\ca$, given in (\ref{4.1a}), as $n \rt \infty$.

The kernels $I_{p_{n}}$ and $I_{q_{n}}$ of $p_n$ and $q_n$ are respectively
\bea
& & I_{p_{n}} = \left\{ \left[
\begin{array}{cc}
m_{n \times n} & 0 \\
0 &0
\end{array}
\right] \right\}~,
\nonumber \\
& & I_{q_n} = \{ 0 \}~.
\label{6.10}
\eea
Since $I_{q_n} \subset I_{p_n}$, the hull kernel topology on the set
$\{ p_n, q_n \}$ is given by $q_n \prec p_n$. Hence $\{p_n ,q_n\}$
is the two-point poset
shown in Fig. 8(a) and  is exactly the same as the one in Fig. 5(a).

\begin{figure}[hbtp]
\begin{center}
\large
\mbox{\psannotate{\psboxto(0cm;6cm){fig6.eps}}{ %\fillinggrid
\at(5\pscm;4\pscm){$q_n$}
\at(4.5\pscm;11\pscm){$p_n$}
\at(19\pscm;4\pscm){$a_n(q_n)=a_n$}
\at(18.5\pscm;11\pscm){$a_n(p_n)=\l$}
}}
\normalsize
\end{center}
{\footnotesize{\bf Fig.8.}
(a) is the poset for the algebra $\ca_n$ of (\ref{6.1}) while
(b) shows the values of a typical element $a_n$ of this algebra at its
two points $p_n$ and $q_n$.}
\end{figure}

Thus the preceding two representations of $\ca_{n}$ form a topological
space identical to the poset of $\ca$.

All this suggests that it is possible to approximate $\ca$ by $\ca_{n}$ and
regard its representations $p_n$ and $q_n$ as constituting the
configuration space.

In our previous discussions, either involving the algebra $\cc$ or
the algebra $\ca$, we considered only their IRR's. But the
representation $q_n$ of $\ca_{n}$ is not IRR. It has the invariant
subspace
\be
{\bf C} ~ \left( \begin{array}{c} 0 \\ 0 \\ \vdots \\ 0 \\ 1
\end{array} \right)  ~ ~ .\label{6.11}
\ee
In this respect we differ from the previous sections in our treatment
of $\ca_{n}$.

The first nontrivial approximation is $\ca_1$.
It is a commutative algebra with elements
\be
\left( \begin{array}{cc} \l_1 & 0 \\ 0 & \l_2 \end{array} \right)
\equiv (\l_1,\l_2) ~ ,~ \l_i \in {\bf C} ~ . \label{6.12}
\ee
In this way, we can achieve a commutative simplification of $\ca$ which
will be denoted by $\cc(\ca)$.

Let us now consider the $\bigvee$ poset and its algebra $\ca =
{\bf C}\cp_1 + \ck_{12}+ {\bf C}\cp_2$ acting on the Hilbert space
$\ch_1 \oplus \ch_2$. Its finite dimensional
approximations are given by
\bea
& & \ca_0 = {\bf C} , \nonumber \\
& & \ca_1 = {\bf C} \oplus {\bf C}~, \nonumber \\
& & \ca_2 = {\bf C} \oplus M(2, {\bf C}) \oplus {\bf C}~, \nonumber \\
& & \cdots \nonumber \\
& & \ca_n = {\bf C} \oplus M(2n-2, {\bf C}) \oplus {\bf C}~, \nonumber
\\ & & \ldots~~~~~~~~~~~~~
\label{6.12.1}
\eea
where a typical element of $a_{n} \in \ca_{n}$ is of the form
\be
a_{n} = \left[
\begin{array}{ccc}
\l_1 & 0              & 0    \\
0    & m_{2n-2 \times 2n-2} & 0    \\
0    & 0              & \l_2
\end{array}
\right]~.
\label{6.12.2}
\ee

As before, there is a *-homomorphism
\be
F_{n+1,n} : \ca_{n} \rightarrow \ca_{n+1} ~ , \label{6.17}
\ee
given by
\be
a_{n} = \left[
\begin{array}{ccc}
\l_1 & 0                  & 0    \\
0    & m_{2n-2 \times 2n-2} & 0    \\
0    & 0                  & \l_2
\end{array}
\right]~~ \rt ~
\left[
\begin{array}{ccccc}
\l_1 & 0    & 0                  & 0       & 0      \\
0    & \l_1 & 0                  & 0       & 0      \\
0    & 0    & m_{2n-2 \times 2n-2} & 0       & 0      \\
0    & 0    & 0                  & \l_2    & 0       \\
0    & 0    & 0                  & 0       & \l_2
\end{array}
\right]~.
\label{6.12.3}
\ee
We thus have a directed system of C$^{*}$-algebras whose inductive limit
is $\ca$ \cite{FD}, showing that $\ca_{n}$ approximates $\ca$.

The algebras $\ca_{n}$ have the following three representations:
\begin{eqnarray}
a) ~~~~~~~~~~ \a_n & : & a_n \rightarrow \l_1 ~ , \nonumber \\
b) ~~~~~~~~~~ \b_n & : & a_n \rightarrow \l_2 ~ , \nonumber \\
c) ~~~~~~~~~~ \g_n & : & a_n \rightarrow a_n ~ . \label{6.18}
\end{eqnarray}
Note that $\a_n$ and $\b_n$ are commutative IRR's while $\g_n$ is not IRR,
just like $q_n$.

Now the kernels of these representations are
\begin{eqnarray}
I_{\a_{n}}& = & \{0\} \oplus M(2n-2, {\bf C}) \oplus {\bf C} ~ , \nonumber \\
I_{\b_{n}}& = & {\bf C} \oplus M(2n-2, {\bf C}) \oplus \{0\} ~ , \nonumber \\
I_{\g_{n}}& = & \{ 0 \} ~ . \label{6.20}
\end{eqnarray}

Since
\be
I_{\g_{n}} \subset I_{\a_{n}} \mbox{ and } I_{\g_{n}} \subset I_{\b_{n}} ~ ,
\label{6.21}
\ee
we set
\be
\g_n \prec \a_n ~ , ~ \g_n \prec \b_n ~ . \label{6.22}
\ee
The poset that results is shown in Fig. 9. It is again the $\bigvee$ poset,
suggesting that $\ca_{n}$ and its representations $\a_n,\b_n,\g_n$ are
good approximations for our purposes.\\

\begin{figure}[hbtp]
\begin{center}
\large
\mbox{\psannotate{\psboxto(0cm;7cm){fig7.eps}}{ %\fillinggrid
\at(6\pscm;4\pscm){$\g_n $}
\at(3\pscm;11\pscm){$\a_n $}
\at(8\pscm;11\pscm){$\b_n $}
\at(17\pscm;4\pscm){$a_n(\g_n)=a_n $}
\at(14.5\pscm;11\pscm){$a_n(\a_n)=\l_1 $}
\at(20\pscm;11\pscm){$a_n(\b_n)=\l_2 $}
}}
\normalsize
\end{center}
{\footnotesize{\bf Fig.9.}
(a) is the poset for the algebra $\ca_{n}$ defined in
(\ref{6.1}) while (b) shows the values of a typical element $a_n$ of
this algebra at its three points $\a_n,\b_n,\g_n$. }
\end{figure}

Now the C$^{*}$-algebra
\be
\ca_{1} = {\bf C} \oplus {\bf C} =
\left\{ a_1 = \left[
\begin{array}{cc}
\l_1 & 0     \\
0    & \l_2
\end{array}
\right]~;~ \l_i \in {\bf C} \right\}
\label{6.23}
\ee
is commutative and its representations $\a_1, \b_1$ and $\g_1$ also capture
the poset topology correctly. It will be denoted again by $\cc(\ca)$.
It seems to be the algebra with the minimum number of degrees of
freedom correctly reproducing the poset and its topology.

Is it possible to interpret $\l_i$?
For this purpose, let us remember that the points of a manifold $M$
are closed, and so correspond to the top or level one points of the poset. The
latter somehow approximate the former. Since the values of $a_1$ at
the level one points are $\l_1$ and $\l_2$, we can regard $\l_i$ as
the values of a continuous function on $M$ when restricted to this
discrete set. The role of the bottom points in the poset and the value of
$a_1$ there is to somehow glue the top points together and generate a
nontrivial approximation to the topology of $M$.

We can explain this interpretation further using simplicial
decomposition. Thus the interval $[0,1]$ has a simplicial
decomposition with $[0]$ and $[1]$ as zero-simplices and $[0,1]$ as
the one-simplex. Assuming that experimenters can not resolve two
points if every simplex containing one contains also the other, they
will regard $[0,1]$ to consist of the three points $\a_1 =[0]$, $\b_1
= [1]$ and $\g_1=]0,1[$. There is also a natural map from
$[0,1]$ to these points as in Section \ref{se:2}. Introducing
the quotient topology on these points following that section,
we get back the $\bigvee$ poset. In this approach then,
$\l_1$ and $\l_2$ are the values of a continuous function at the two
extreme points of $[0,1]$ whereas the association of $\l_1 \cp_1 +
\l_2 \cp_2$ with the open interval is necessary to cement the extreme
points together in a topologically correct manner.

[We remark here that the simplicial decomposition of any manifold
yields a poset in the manner just indicated. We will also suggest
at the  end of Section \ref{se:4.2} that a probability density can not be
localized at level one points, unless they are isolated and hence both open and
closed. This result appears eminently
reasonable in the context of a simplicial decomposition where level
one points are points of the manifold. Reasoning like this also
suggests that localization must in general be possible only at the subsets of
the poset representing the open sets of $M$. That seems in fact to be the
case. For, as will be indicated in Section \ref{se:4.2}, localization seems
possible only at the open sets of a poset and the latter
correspond to open sets of $M$.]

\sxn{Quantum Theory Using $\ca$}\label{se:4.2}

The noncommutative algebra $\ca$ is an algebra of observables. It
replaces the algebra $\cc(M)$ when $M$ is approximated by a poset. We
must now find the space $\ce$ on which $\ca$ acts, convert $\ce$ into a
pre-Hilbert space and therefrom get the Hilbert space $\ch$ by
completion.

Now as $\ca$ is noncommutative, it turns out to be important to specify
if $\ca$ acts on $\ce$ from the right or the left. We will take the action
of $\ca$ on $\ce$ to be from the right, thereby making $\ce$ a right
$\ca$-module.

The simplest model for $\ce$ is obtained from $\ca$ itself. As for the
scalar product, note that $(\x^* \h)(x)$ is an operator in a Hilbert space
$\ch_x$ if $\x, ~ \h \in \ca$ and $x \in$ poset. We can hence find a
scalar product $(\cdot ,\cdot )$
by first taking its operator trace $Tr$ on $\ch_x$ and then summing it
over $\ch_x$ with suitable weights $\r_x$:
\be
(\x,\h) = \sum_x \r_x Tr (\x^* \h) (x), ~~~ \r_x \geq 0~. \label{4.4}
\ee

As remarked in Section \ref{se:3}, there is no need for $\ce$ to be $\ca$. It
can be any space with the following properties:
\begin{enumerate}
\item[(1)] It is a right $\ca$-module. So, if $\x \in \ce$ and $a \in
\ca$, then $\x a \in \ce$.

\item[(2)] There is a positive definite ``sesquilinear"
form $<\cdot , \cdot>$ on $\ce$ with values in
$\ca$. That is, if $\x, \h \in \ce$, and $a \in \ca$, then
\item[a)]
\bea
& & <\x, \h> \in \ca~,~~~  <\x, \h> ^\ast =  <\h, \x>~, \nonumber \\
& & <\x, \x> \geq 0 ~~{\rm and}~~  <\x, \x> = 0 \iff \x = 0~.
\eea
\end{enumerate}
\noindent
Here $``<\x, \x > \geq 0"$ means that it can be written as $a^* a$ for
some $a \in \ca$.
\begin{enumerate}
\item[b)]
\be
<\x, \h a> = <\x, \h>a~,~~~ <\x a, \h> = a^* <\x, \h>~.
\ee
\end{enumerate}

The scalar product is then given by
\be
(\x,\h) = \sum_x \r_x Tr  <\x, \h> (x)~. \label{4.5}
\ee

As $\x^* \h(x), <\x, \h>(x), <\x, \h a>(x)$ or $<a\x, \h>(x)$ may not be
of trace class \cite{Si}, there are questions of convergence
associated with (\ref{4.4}) and (\ref{4.5}). We presume that these
traces must be judiciously regularized and modified
(using for example the Dixmier
trace \cite{Co,VG}) or suitable conditions put on
$\ce$ or both.
But we will not address
such questions in detail in this article.

When $\ca$ is commutative and has structure space $M$, then an $\ce$ with
the properties described consists of sections of hermitian vector
bundles over $M$. Thus, the above definition of $\ce$ achieves a
generalization of the familiar notion of sections of hermitian
vector bundles to noncommutative geometry.

In the literature
\cite{Co,VG}, a method is available for the
algebraic construction of $\ce$. It works both when $\ca$ is
commutative and noncommutative. In the former case, Serre and Swan
\cite{Co,VG} also
prove that this construction gives (essentially) all $\ce$ of physical
interest, namely all $\ce$ consisting of sections of vector bundles.
It is as follows.
Consider $\ca \otimes {\bf C}^N \equiv \ca^N$ for some integer $N$.
This space consists of $N$-dimensional vectors with coefficients in $\ca$
(that is, with elements of $\ca$ as entries). We can act on it from
the left with $N \times N$ matrices with coefficients in $\ca$.
Let $ e = [e^i_j]$ be such a matrix which is idempotent, $e^2 = e$, and
hermitian, $< e \x , \h> = <\x ,e \h >$.
Then, $ e \ca^N$ is an $\ce$, and according to the Serre-Swan theorem,
every $\ce$ [in the sense above]
is given by this expression for some $N$ and some $e$ for commutative $\ca $.
An $\ce$ of the form $e \ca^N$ is called a ``projective module
of finite type" or a ``finite projective module".

Note that such $\ce$
are right $\ca$-modules. For, if $\x \in e \ca^N$, it
can be written as a vector $(\x^1, \x^2, \cdots, \x^N)$ with $\x^i
\in \ca$ and $e_j^i \x^j = \x^i$. The action of $a \in \ca$ on $\ce$ is
\be
\x \rt \x a = (\x^1 a, \x^2 a, \cdots, \x^N a)~. \label{4.6}
\ee

With this formula for $\ce$, it is readily seen that there are many
choices for $< \cdot~,~\cdot >$~. Thus let $g = [g_{ij}]$, $g_{ij} \in \ca$,
be an $N \times N$ matrix with the following properties: a) $g_{ij}^* =
g_{ji}$;
b) $\x^{i*} g_{ij} \x^j \geq 0$ and $\x^{i*} g_{ij} \x^j = 0 \Leftrightarrow
\x=0$.
Then, if $\h = (\h^1, \h^2, \cdots, \h^N)$ is another vector in $\ce$, we
can set
\be
<\x, \h> = \x^{i*}  g_{ij} \h^j~. \label{4.7}
\ee

In connection with (\ref{4.7}), note that the algebras $\ca$ we consider here
generally have unity. In those cases, the choice $g_{ij} \in {\bf C}$ is a
special case of the condition $g_{ij} \in \ca$. But if $\ca$ has no unity, we
should also allow the choice $g_{ij} \in {\bf C}$ .

The minimum we need for quantum theory is a Laplacian $\D$ and
a potential function $W$, as a Hamiltonian can be constructed from
these ingredients. We now outline how to write $\D$ and $W$.

Let us first look at $\D$, and assume in the first instance that
$\ce = \ca$.

An element $a \in \ca$ defines the operator $\oplus_x a(x)$ on the
Hilbert space $\ch = \oplus_x \ch_x$, the map $a \rt \oplus_x a(x)$ giving a
faithful representation of $\ca$. So let us identify $a$ with
$\oplus_x a(x)$ and $\ca$ with this representation of $\ca$ for the
present.

In noncommutative geometry \cite{Co,VG}, $\D$ is constructed from an
operator $D$ with specific properties on
$\ch$. The operator $D$ must be self-adjoint and the commutator
$[D, a]$ must be bounded for all $a \in \ca$:
\be
D^* = D~,~~~ [D, a]  \in \cb~~~~\mbox{for all}~~ a \in \ca~.
\label{4.7.1}
\ee
Given $D$, we construct the `exterior derivative' of
any $a \in \ca$ by setting
\be
da = [D, a] := [D, \oplus_x a_x] .
\label{4.8}
\ee
Note that $da$ need not be in $\ca$, but it is in $\cb$.

Next we introduce a scalar product on $\cb$ by setting
\be
(\a, \b) = Tr [\a^{*} \b]~,~~~\mbox{for all}~~ \a, \b \in \cb~, \label{4.9}
\ee
the trace being in $\ch$.
[Restricted to $\ca$, it becomes (\ref{4.5}) with $\r_x = 1$. This
choice of $\r_x$ is made for simplicity and can readily dispensed
with. See also the comment after (\ref{4.5})]

Let $p$ be the orthogonal projection operator on $\ca$ for this
scalar product:
\bea
& & p^2 = p^* = p~, \nonumber \\
& & p a = a ~~~\mbox{if}~~~ a \in \ca~, \nonumber \\
& & p \a = 0 ~~~\mbox{if}~~~ (a, \a)=0~~~\mbox{and}~~~
a \in \ca~. \label{4.9.1}
\eea

The Laplacian $\D$ on $\ca$ is defined using $p$ as follows.

We first introduce the adjoint $\d$ of $d$. It is an operator from $\cb$ to
$\ca$:
\be
\d~:~\cb \rightarrow \ca~.\label{4.9a}
\ee
It is defined as follows. Consider all $b\in \cb$ for which $(b,da)$ can be
written as $(a',a)$ for all $a\in \ca$. Here $a'$ is an element of $\ca$ linear
in the elements $b$ and independent of $a$. Thus
\be
(b, d a) = (a', a) ~, ~~ \forall~ a \mbox{ and some }
a'\in \ca~. \label{4.10}
\ee
Then we write
\be
a'= \d b~~.\label{4.10a}
\ee
A computation shows that
\be
\d b=p[D,b]~~.\label{4.10b}
\ee
The Laplacian can now be defined as usual as
\be
\D a \equiv - \d d a = - p [D, [D, a]]~. \label{4.11}
\ee
Notice that the domain of $\D$ does not necessarily coincide with $\ca$.

As for $W$, it is essentially any element of $\ca$. (There may be
restrictions on $W$ from positivity requirements on the Hamiltonian.)
It acts on a wave function $a$ according to $a \rt aW$,
where $(aW)(x) = a(x)W(x)$~.

A possible Hamiltonian $H$ now is $- \l \D + W~,~ \l > 0$~, while a
Schr\"odinger equation is
\be
i \frac{\pa a}{\pa t} = - \l \D a + a W~. \label{4.12}
\ee

When $\ce$ is a nontrivial projective module of finite type over
$\ca$, it is necessary to introduce a connection and ``lift" $d$
from $\ca$ to an operator $\del$ on $\ce$.
Let us assume that $\ce$ is obtained from the construction
described before (\ref{4.6}).
In that case the definition of $\del$ proceeds as follows.

Because of our assumption, an element $\x \in \ce$ is given by
$\x = (\x^1, \x^2, \cdots, \x^N)$ where $\x^i \in \ca$ and $e^i_j
\x^j = \x^i$. Thus $\ce$ is a subspace of $\ca \otimes {\bf C}^N :=
\ca^N$:
\bea
& & \ce \subseteq \ca^N~, \nonumber \\
& & \ca^N = \{ (a^1, \cdots, a^N)~:~ a^i \in \ca \}~. \label{4.13.1}
\eea
Here we regard $a^i$ as operators on $\ch$. Now $\ca^N$ is a subspace
of $\cb \otimes {\bf C}^N := \cb^N$ where $\cb$ consists of bounded
operators on $\ch$. Thus
\bea
& & \ce \subseteq \ca^N \subseteq \cb^N~, \nonumber \\
& & \cb^N = \{ (\a^1, \cdots, \a^N)~:~ \a^i =
{}~\mbox{bounded operator on}~ \ch \}~. \label{4.13.2}
\eea

Let us extend the scalar product $(\cdot, \cdot)$ on $\ce$ [given by
(\ref{4.7}) and (\ref{4.5})] to $\cb^N$ by setting
\bea
& & < \a, \b > = \a^{i*} g_{ij} \b^j~, \nonumber \\
& & (\a, \b) = Tr <\a, \b> ~~~\mbox{for}~~ \a, \b \in \cb^N~.
\label{4.13.3}
\eea

Next, having fixed $d$ on $\ca$ by a choice of $D$ as in
(\ref{4.7.1}), we define $d$ on $\ce$ by
\be
d \x = (d\x^1, d\x^2, \cdots, d\x^N)~. \label{4.13.4}
\ee
Note that $d\x$ may not be in $\ce$, but it is in $\cb^N$:
\be
d \x \in \cb^N~. \label{4.15.5}
\ee

A possible $\nabla$ for this $d$ is
\be
\nabla \x = e d \x + \r \x~, \label{4.15.6}
\ee
where
\begin{enumerate}
\item[{\rm a)}] $e$ is the matrix introduced earlier,
\item[{\rm b)}] $\r$ is an $N \times N$ matrix with coefficients in $\cb$ :
\be
\r = [\r^i_j]~, ~~\r^i_j \in \cb~, \label{4.15.7}
\ee
\item[{\rm c)}]
\be
\r = e \r e~, \label{4.15.8}
\ee
\item[{\rm and}]
\item[{\rm d)}] $\r$ is hermitian:
\be
< \a, \r \b> - < \r \a, \b>= 0~. \label{4.15.9}
\ee
\end{enumerate}

Note that if $\hat{\r}$ fulfills all conditions but $c)$, then $\r =
e \hat{\r} e$ fulfills $c)$ as well.

Condition (\ref{4.15.9}) is equivalent to the compatibility of $\nabla$ with
the hermitian structure (\ref{4.13.3}):
\be
d<\a , \b> =< \a, \nabla \b> - < \nabla \a, \b> ~. \label{4.15.10}
\ee

Having chosen a $\del$, we can try defining  $\del^{*} \del$ using
\be
(\del \x, \del \h) = (\x, \del^{*} \del\h )~, ~~~ \x, \h \in
\ce \label{4.16}
\ee
where $( \cdot, \cdot )$ is defined by (\ref{4.13.3}).
A calculation similar to the one done to define the Laplacian
(\ref{4.11}) on $\ca$ then shows that we can define $\D$ on $\ce$ by
\be
\D \eta = - q \del ^{*} \del \eta ~,~~~ \h \in \ce~, \label{4.17}
\ee
where $q$ is the orthogonal projector on $\ce$ for the scalar product
$( \cdot, \cdot )$.

A potential $W$ is an element of $\ca$. It acts on $\ce$ according to
the rule (1) following (\ref{4.4}).

A Hamiltonian as before has the form $ - \l \D + W~,~ \l > 0$~.
It gives the Schr\"odinger equation
\be
i \frac{\pa \x}{\pa t} = -
\l \D \x + \x W ~ , \x \in \ce ~ . \label{4.15}
\ee

We will not try to find explicit examples for $\D$ here. That task
will be taken up for a simple problem in Section \ref{se:7}.

We will conclude this section by pointing out an interesting property
of states for posets. It does not seem possible to localize a state at the
level one points [unless they happen to be isolated points, both open
and closed]. We can see this for example from Fig. 6(b) which shows
that if a probability density vanishes at $\g$, then $\l_i$ (and
$k_{12}$)
are zero and therefore they vanish also at $\a$ and $\b$. It seems possible
to  show
in a similar way that localization in an arbitrary poset is possible
only at open sets.

\sxn{Line Bundles on Circle Poset and $\q$-quantisation }   \label{se:7}

A circle $S^1 = \{e^{i\f}\}$ is an infinitely connected space. It has
the fundamental group ${\bf Z}$. Its universal covering
space \cite{BMSS} is the real line
${\bf R}^1 = \{ x : -\infty < x < \infty \}$.
The fundamental group ${\bf Z}$ acts on ${\bf R}^1$ according to
\be
x \rightarrow x +N ~ , ~ N \in {\bf Z} ~ . \label{7.1}
\ee
The quotient of ${\bf R}^1$ by this action is $S^1$, the projection
map ${\bf R}^1 \rightarrow S^1$ being
\be
x \rightarrow e^{i2\p x} ~ . \label{7.2}
\ee

Now the domain of a typical Hamiltonian for a particle on $S^1$ need
not consist of smooth functions on $S^1$. Rather it can be obtained from
functions $\psi_{\theta}$ on ${\bf R}^1$ transforming by an IRR
\be
\r_{\q} : N \rightarrow e^{iN\q} \label{7.3}
\ee
of ${\bf Z}$ according to
\be
\psi_{\q}(x+N) = e^{iN\q} \psi_{\q}(x) ~ . \label{7.4}
\ee
The domain $D_{\q}(H)$ for a typical Hamiltonian $H$ then consists of
these $\psi_{\q}$ restricted to a fundamental domain $0 \leq x \leq
1$ for the action
of ${\bf Z}$ and subjected to a differentiability requirement:
\be
D_{\q}(H) = \{ \psi_{\q} : \psi_{\q}(1) = e^{i\q} \psi_{\q}(0)~;~~
\frac{d\psi_{\q}(1)}{dx} = e^{i\q} \frac{d\psi_{\q}(0)}{dx} \} ~
. \label{7.5}
\ee
In addition, of course, if $dx$ is the measure on $S^1$ used to define
the scalar product of wave functions, then $H\psi_{\q}$ must be square
integrable for this measure. It is also assumed that $\psi_{\q}$ is
suitably smooth in $]0, 1[$.

We obtain a distinct quantisation, called $\q$-quantisation, for
each choice of $e^{i\q}$.

As has been shown earlier \cite{BBET}, there are similar quantisation
possibilities for a circle poset as well. The fundamental group of a
circle poset is ${\bf Z}$. Its universal covering space is the poset
of Fig. 10. Its quotient, for example by the action
\begin{eqnarray}
N   & : & x_j \rightarrow x_{j+3N} ~ , \nonumber \\
x_j & = & a_j \mbox{ or } b_j ~ \mbox{ of Fig. 10 }, ~ N \in {\bf Z}
\label{7.6}
\end{eqnarray}
gives the circle poset of Fig. 7(b).

\begin{figure}[hbtp]
\begin{center}
\large
\mbox{\psannotate{\psboxto(0cm;5cm){fig13.eps}}{ %\fillinggrid
\at(5.5\pscm;1\pscm){$b_{-2} $}
\at(8.8\pscm;1\pscm){$b_{-1} $}
\at(12\pscm;1\pscm){$b_0 $}
\at(15.4\pscm;1\pscm){$b_1 $}
\at(18.5\pscm;1\pscm){$b_2 $}
\at(6\pscm;6\pscm){$a_{-2} $}
\at(9.5\pscm;6\pscm){$a_{-1} $}
\at(12.5\pscm;6\pscm){$a_0 $}
\at(16\pscm;6\pscm){$a_1 $}
\at(19\pscm;6\pscm){$a_2 $}
}}
\normalsize
\end{center}
\begin{center}
{\footnotesize{\bf Fig.10.}
The figure shows the universal covering space of a circle
poset.}
\end{center}
\end{figure}

In \cite{BBET}, it has been argued that the poset analogue
of $\q$-quantisation can be obtained from complex
functions $f$ on the poset of Fig. 10 transforming by an IRR of
${\bf Z}$:
\be
f(x_{j +3}) = e^{i\q} f(x_j) ~ .\label{7.7}
\ee
While answers such as the spectrum of a typical Hamiltonian came out
correctly, this approach was nevertheless
affected by a serious defect: continuous complex functions on a
connected poset are constants, so that our wave functions can not be
regarded as continuous.

This defect was subsequently repaired in \cite{lisbon}
by using the algebra $\cc(\ca)$
for a circle poset and the corresponding algebra $\bar{\cc}(\ca)$ for
Fig. 10.

In this article we give an alternative description of the latter
approach to quantisation.
We shall construct, much in the spirit of Section \ref{se:4.2},
the algebraic analogue of the trivial
bundle on the poset $P_N(S^1)$ with a `gauge connection' such that the
corresponding Laplacian gives the answer of ref. \cite{lisbon}.

The algebra $\cc(\ca)$ associated with the poset  $P_{2N}(S^1)$ is given by
\be
\cc(\ca) = \{ c=(\l_1 ,\l_2 ) \oplus (\l_2 ,\l_3 ) \oplus \cdots
\oplus (\l_N ,\l_1 )~
: ~ \l_i \in {{\bf C}} \} ~ . \label{7.8}
\ee

The ``finite projective module of sections" $\ce$ associated with the
trivial bundle is taken to be $\cc(\ca)$ itself so that the $e$ of
Section \ref{se:4.2} is the identity.
To avoid confusion between the dual roles $\cc(\ca)$ and $\ce$ of
the same set, we indicate the elements of $\ce$ using the letter $\m$
[whereas we use $\l$ for those of $\cc(\ca)$] :
\begin{eqnarray}
\ce = \{ \c ,\c ' ,\ldots & : &
\c = (\m_1 ,\m_2 ) \oplus (\m_2 ,\m_3 )
\oplus \cdots \oplus (\m_N , \m_1 )~, \nonumber \\
& & \c '= (\m'_1 , \m'_2 ) \oplus (\m'_2 ,\m'_3 )
\oplus \cdots \oplus (\m'_N , \m'_1 ) ~ , ~ \ldots
{}~ ; ~ \m_i , \m'_i \in {\bf C} \} ~ . \label{7.11}
\end{eqnarray}

Here
$\ce$ is a $\cc(\ca)$-module, with the action of $c$ on $\c$
given by
\be
\c c = (\m_1 \l_1,\m_2 \l_2 ) \oplus (\m_2 \l_2,\m_3 \l_3)
\oplus \cdots (\m_N \l_N, \m_1 \l_1) ~ . \label{7.12}
\ee

The space $\ce$ has a sesquilinear form $\langle \cdot,\cdot \rangle$
valued in $\cc(\ca)$:
\be
\langle \c' , \c \rangle :=
(\m_1^{'*} \m_1,\m_2^{'*} \m_2 ) \oplus (\m_2^{'*} \m_2,\m_3^{'*} \m_3)
\oplus \cdots \oplus (\m_N^{'*} \m_N, \m_1^{'*} \m_1) ~ \in \cc(\ca)~ .
\label{7.13}
\ee

An equivalent realization of $\cc(\ca)$ (and hence of $\ce$) can be
given in terms of $N \times N$ diagonal matrices, typical elements of
$\cc(\ca)$ and $\ce$ in this new realization being
\bea
&& \overline{c} = \mbox{diag}(\l_1, \l_2, \dots \l_N)~, \nonumber \\
&& \overline{\c} = \mbox{diag}(\m_1, \m_2, \dots \m_N)~. \label{7.14.1}
\eea
The scalar product associated with (\ref{7.13}) can be written, after
a rescaling, as
\be
(\overline{\c}',\overline{\c}) = \sum_{j=1}^N \m'^*_j \m_j
= \mbox{Tr} \overline{\c}'^* \overline{\c}
{}~. \label{7.14.2}
\ee

% Start of corrections 28 feb 95

In order to define a Laplacian, we need an operator $D$ like in
(\ref{4.7.1}) to define the `exterior derivative' $d$ of (\ref{4.8}), and a
matrix of one-forms $\r$ with the properties (\ref{4.15.7})-(\ref{4.15.9})
which is the analogue of the connection form.
Assuming the identification of  $N + j$ with $j$,
we take for $D$ the self-adjoint matrix with elements
\be
D_{ij} = \frac{1}{\sqrt{2}\epsilon} (m^* \d_{i+1,j} + m \d_{i,j+1})~,
{}~ i,j = 1, \cdots, N~, \label{7.14.3}
\ee
where $m$ is any complex number of modulus one, $m m^* =1$.
As for the connection $\r$, we take it to be the hermitian matrix with
elements
\bea
&&\r_{ij} =  \frac{1}{\sqrt{2}\epsilon}
(\s^* m^* \d_{i+1,j} + \s m \d_{i,j+1})~, \nonumber \\
&&~~~ \s = e^{- i \q / N} - 1~,
{}~~i,j = 1, \cdots, N~.
\label{7.14.4}
\eea
One checks that the curvature of $\r$ vanishes, namely
\footnote{A better analysis should take into account the
structure of the `junk forms' \cite{Co,VG}.
This can be done, but we do not give details here.}
\be
d \r + \r^2 = 0~. \label{7.14.5}
\ee
It is also possible to prove that $\r$ is a `pure gauge', that is
that there exists a $\overline{c} \in \cc(\ca)$ such that
$\r = \overline{c}^{-1} d \overline{c}$, only for $\theta = 2 \p k,
k$~ any integer.
[If $\overline{c}=\mbox{diag}(\l_1 ,\l_2, \ldots ,\l_N )$,
any such $\overline{c}$ will be given by $\l_1 = \l~,
{}~\l_2 = e^{i 2\p k / N } \l~, ... ,
{}~\l_j = e^{i 2\p k (j-1) / N } \l~, ...,
{}~\l_N = e^{i 2\p k (N - 1) / N } \l~, \l$ not equal to $0$.]
These properties are the analogues of the
the well-known properties of the
connection for a particle on $S^1$ subjected to $\q$-quantisation,
with single-valued wave functions on $S^1$ defining the domain of the
Hamiltonian. If the Hamiltonian with the domain (\ref{7.5}) is
$-d^2 / d x^2$, then the Hamiltonian with the domain $D_0(h)$
consisting of single valued wave functions is $-(d / d x + i \q)^2$
while the connection one-form is $i \q d x$.

We next define $\nabla$ on $\ce$ by
\be
\nabla \overline{\c} = [D, \overline{\c}] + \r \overline{\c}~,
\label{7.14.5.1}
\ee
in accordance with (\ref{4.15.6}), $e$ being the identity.
The covariant Laplacian $\D_\q$ can then be computed as follows.

% end of corrections 28 feb 95

We write
\be
(\del \overline{\c}', \del \overline{\c}) =
(\overline{\c}', \del^{*} \del\overline{\c} )~, \label{7.14.6}
\ee
as in (\ref{4.16}). Now the projection operator $q$ in the present
case is readly seen to be defined by
\be
(q M)_{ij} = M_{ii}\delta_{ij}~,~~~\mbox{no summation on}~i~,
\label{7.14.7}
\ee
$M$ being any $N \times N$ matrix. Hence
\bea
 (\D_\q \overline{\c})_{ij} &=& - (\del^{*} \del\overline{\c})_{ii}
\delta_{ij}~, \nonumber \\
- (\del^{*} \del\overline{\c})_{ii} &=&
\left\{ - \left[ D, [D, \overline{\c} ] \right] - 2 \r [D,
\overline{\c}] - \r^2 \overline{\c} \right\}_{ii}
= \frac{1}{\epsilon^2}
\left[ e^{-i\q/N} \m_{i-1} - 2 \m_i +  e^{i\q/N} \m_{i+1} \right]~;
\nonumber \\
i &=& 1, 2, \cdots , N ;~~ \m_{N+1} = \m_1~.
\label{7.15}
\eea

The solutions of the eigenvalue problem
\be
\D_\q \overline{\c} = \l \overline{\c} \label{7.16}
\ee
are
\bea
\l &=& \l_k = \frac{2}{\epsilon ^2}
\left[\cos (k + \frac{\theta}{N}) -1 \right]~, \nonumber \\
\overline{\c} &=& \overline{\c}^{(k)}
= \mbox{diag}(\m^{(k)}_1, \m^{(k)}_2, \cdots, \m^{(k)}_N)~,
{}~~~k = m \frac{2\p}{N}~,~ m = 1, 2, \cdots, N~, \label{7.16.1}
\eea
where
\be
\m^{(k)}_j = A^{(k)} e^{i kj} + B^{(k)}e^{-i kj}~,~~~
A^{(k)}, B^{(k)}\in {\bf C}~. \label{7.17}
\ee

These are exactly the answers in ref. \cite{BBET} but for one
significant difference. In ref. \cite{BBET}, the operator $\D$ did not
mix the values of the wave function at points of level one and level
two, resulting in a double degeneracy of eigenvalues. That unphysical
degeneracy has now been removed because of a better treatment of
continuity properties. The latter prevents us from giving independent values
to continuous probability densities at these two
kinds of points. Note that the approach of ref. \cite{lisbon}
is equivalent to the present one and also give (\ref{7.17}) without the
spurious degeneracy.

\sxn{Abelianization and Gauge Invariance} \label{se:5}

\subsxn{Commutative Subalgebras and Unitary Groups}\label{se:5.1}

As remarked in Section \ref{se:6},
the \cstars for our posets are
approximately finite dimensional (AF) \cite{Br,SV}.
Besides the ones described there,
they have additional nice structural properties
which can be exploited to
develop relatively transparent models for $\ce$.
Furthermore, these properties are of use in the analysis of the limit
where the number of points of the poset approximation is allowed to go to
infinity. This will be explored in a future publication.

Here we will describe very simple and physically suggestive
presentations of such algebras in terms of their
maximal commutative subalgebras. We will then use this presentation to derive
the commutative algebra $\cc(\ca)$ using a gauge principle.

We will start with some definitions \cite{Br,SV}.
The commutant $A'$ of a
subalgebra $A$ of $\ca$ consists of all elements of $\ca$ commuting
with all elements of $A$ :
\be
A' = \left \{ x \in \ca : xy = yx~, ~~\forall~ y \in A \right \}~.
\label{5.1}
\ee

A maximal commutative subalgebra $\cc$ of $\ca$ is a commutative
C$^*$-subalgebra of $\ca$ which coincides
with its commutant, $\cc ' = \cc$.

The C$^*$-algebras $\ca$ we consider have a unity
$1\!\!1$. We therefore have the concepts of the inverse and unitary
elements for $\ca$.

Let $\cc$ be a maximal commutative subalgebra of $\ca$ and let $\cu$ be
the normalizer of $\cc$ among the unitary elements of $\ca$:
\be
\cu = \left \{ u \in \ca ~|~ u^* u = 1\!\!1 ~;~ u^* c u \in \cc
{}~~{\rm if}~~ c \in \cc \right \}~.
\label{5.2}
\ee
One can show \cite{SV} that if $u \in \cu$, then
$u^* \in \cu$, so that $\cu$ is a unitary group.

For an AF algebra $\ca$,
a fundamental result in \cite{SV} states that
the algebra generated by $\cc$ and $\cu$ coincides with $\ca$.
If $M_1, M_2, \dots, $ are subsets
of the C$^*$-algebra $\ca$, and we denote by $<M_1, M_2, \dots, >$
the smallest C$^*$-subalgebra of $\ca$
containing $ \bigcup_n M_n$, then the above result can be written as
\be
\ca = < \cc,~ \cu >~.
\label{5.3}
\ee

Next note that $\cc$ in general has unitary
elements and hence $\cu \bigcap \cc \not= \emptyset$. Now $\cu \bigcap
\cc$ is a normal subgroup of $\cu$. We can in fact write $\cu$ as the
semidirect product $\sdp $
of the group $\cu \bigcap \cc$ with a group $U$
isomorphic to $\cu / [\cu \bigcap \cc]$:
\be
\cu = [\cu \bigcap \cc]\sdp U~.
\label{5.7}
\ee
Hence, by (\ref{5.3}),
\be
\ca = < \cc,~ U >~.
\label{5.8}
\ee
This result is of great interest for us.

The group $U$ can be explicitly constructed in cases of interest to us.
We will do so below for the two-point and $\bigvee$ posets. The general
result for any two-level poset follows easily therefrom.

We will now see how $\ca$ can be realized as operators on a suitable
Hilbert space.

Let $\Hat{\cc}$ be the space of IRR's or the structure space of $\cc$. Since
the latter is a commutative AF algebra, we can assert from known results
\cite{Br} that the space $\Hat{\cc}$ is a countable totally disconnected
Hausdorff space, that is, the connected component of each point consists of the
point itself.

If, in the spirit of the Gel'fand-Naimark theorem, we regard elements of $\cc$
as functions on $\Hat{\cc}$, each $x \in \Hat{\cc}$
defines an ideal $I_x$ of $\cc$~:
\be
I_x = \{ f \in \cc ~|~ f(x) = 0 \}~.
\label{5.5}
\ee
Such ideals are called primitive ideals.
They have the following properties for the commutative algebra $\cc$: a)
Every ideal is contained in a primitive ideal, and a primitive ideal is
maximal, that is it is contained in no other ideal; b) A primitive ideal
$I$ uniquely fixes a point $x$ of $\Hat{\cc}$ by the requirement $I_x
=I$. Thus $\Hat{\cc}$ can be identified with the space
Prim$(\Hat{\cc}$) of primitive ideals.

Now if $I_x \in \mbox{Prim}(\Hat{\cc})$ and $c \in \cc$, then $cu^* I_x u =
u^* [ucu^*] I_x u = u^* I_x u$~ since ~$u c u^* \in I_x$.
Similarly  $u^* I_x u c = u^* I_x u$.
Hence $u^* I_x u$ is an ideal.
That being so, there is a primitive ideal $I_y$
containing $u^* I_x u$, $u^* I_x u \subseteq I_y$. Hence $I_x
\subseteq u I_y u^*$. Since $u I_y u^*$ is an ideal too, we conclude
that $I_x = u I_y u^*$ or $u^* I_x u = I_y$.
Calling
\be
y := u^* x = u^{-1} x~,
\label{5.6}
\ee
we thus get an action of $\cu$ on $\Hat{\cc}$. With respect to the
decomposition (\ref{5.7}), only the elements in $U$ act not trivially on
$\Hat{\cc}$, whereas elements in $\cu \cap \cc$ act as the identity.

Let $\ell^2(\Hat{\cc})$ be the Hilbert space of square summable
functions on $\Hat{\cc}$:
\be
(g, h) = \sum_x g(x)^* h(x) < \infty ~,~~~\forall g, h \in
\ell^2(\Hat{\cc})~.
\label{5.4}
\ee

It is a striking theorem of \cite{SV} that $\ca$ can be realized as
operators on
$\ell^2(\Hat{\cc})$ using the formul{\ae}~
\bea
& & (h \cdot f)(x) = h(x)f(x)~, \nonumber \\
& & (h \cdot u)(x) = h(u^* x)~, ~~~ \forall f \in \cc~, u \in U~, h \in
\ell^2(\Hat{\cc})~.
\label{5.9}
\eea
We have shown the action as multiplication on the right in order to be
consistent with the convention in Section \ref{se:4.2}. Also the dot
has been introduced in writing this action for a reason which will
immediately become apparent.

This realization of $\ca$ can give us simple models for $\ce$. To see
this, first note that we had previously
used $\ca$ or $e \ca^n$ as
models for $\ce$. But as elements of $\cc$ are functions on
$\Hat{\cc}$ just like $h$, we now discover that they are also
$\ca$-modules in view of (\ref{5.9}), the relation between the dot
product of (\ref{5.9}) and the algebra product (devoid of the dot)
being
\bea
c \cdot f & = & c f \; , \nonumber \\
c \cdot u & = & u c u^{-1} \; . \label{5.9a}
\eea
The verification of (\ref{5.9a}) is easy.

Thus $\cc$ itself can serve as a simple model for $\ce$.

We may be able to go further along this line since certain
finite projective modules over $\cc$ may also serve as $\ce$.
Recall for this purpose that such a module
is $E \cc^N$ where $E$ is
an $N \times N$ matrix with coefficients in $\cc$, which is idempotent
and hermitian [$E^2 = E~, E^{*} = E~, ~~{\rm where}~~
(E^{*})^i_j = (E^{j}_i)^* $].
A vector in
this module is $\xi = (\xi^1, \xi^2, \cdots, \xi^N)$ with $\xi^i =
E^i_j a^j~,~ a^j \in \cc$.
Now consider
the action $\xi \rightarrow \xi \cdot u$ where $(\xi \cdot u)^i = u
(E^i_j a^j ) u^{-1}$. The vector $\xi \cdot u$ remains in $E \cc^N$
if
\be
u E^i_j u^{-1} = E^i_j \; , \mbox{ that is }, \; u E u^{-1} = E \; .
\label{5.9b}
\ee
Since $\cc$ anyway acts on $E \cc^N$, we get an action of $\ca$
on $E \cc^N$ when (\ref{5.9b}) is fulfilled. Thus $E \cc^N$ is a
model for $\ce$ when $E$ satisfies (\ref{5.9b}).

The scalar product for $\ell^2(\Hat{\cc})$ written above may not be
the most appropriate one and may require modifications or
regularization as we shall see in Section \ref{se:5.2}.
We only mention that
the problem will arise with (\ref{5.4}) because elements of $\ce$ must
belong to $\ell^2(\Hat{\cc})$, a restriction which may be too strong
to give an interesting $\ce$ from $\cc$ or an interesting finite
projective module thereon.

\subsxn{The Two-Point Poset}\label{se:5.2}

We will illustrate the implementation of these ideas for the
two-point, the $\bigvee$ and finally for any two-level poset. That
should be enough to see how to use them for a general poset.

We will treat the two-point poset first. Its algebra is (\ref{4.1}). In
its self-representation $q$, it acts on a Hilbert space $\ch
(=\ch_q)$. Choose an orthonormal basis $h_n~(n = 1, 2, \cdots, )$
for $\ch$ and let $\cp_n$ be the orthogonal projector operator on
${\bf C}h_n$. The maximal commutative subalgebra is then
\be
\cc = < 1\!\!1 ,~ \bigcup_n \cp_n >~.
\label{5.10}
\ee
The structure space of $\cc$ is
\be
\Hat{\cc} = \{ 1, 2, \cdots ; \infty \}~,
\label{5.11}
\ee
where
\bea
a)~~~ n : & & 1\!\!1 \rt 1 := 1\!\!1 (n)~, \nonumber \\
          & & \cp_m \rt \d_{mn} := \cp_m(n)~;
\label{5.12}
\eea
\bea
b)~~~ \infty : & & 1\!\!1 \rt 1 := 1\!\!1 (\infty)~, \nonumber \\
          & & \cp_m \rt 0 := \cp_m(\infty)~.
\label{5.13}
\eea

The topology of $\Hat{\cc}$ is the one given by the one-point
compactification of $\{1, 2, \cdots \}$ by adding $\infty$. A basis
of open sets for this topology is
\bea
& & \{n\}~ ;~ n = 1, 2, \cdots ~~; \nonumber \\
& & \co_k = \{ m ~|~ m \geq k \} \bigcup \{\infty\}~.
\label{5.14}
\eea
A particular consequence of this topology is that the sequence $1, 2,
\cdots, $ converges to $\infty$~.

This topology is identical to the hull kernel
topology \cite{FD}. Thus for instance, the zeros of $\cp_n$ and
$1\!\!1 - \sum_{i=1}^{k-1} \cp_i$ are $\{1, 2, \cdots, \hat{n}, n+1,
\cdots, \infty \}$ and $\{1, 2, \cdots, k-1\}$, respectively, where
the hatted entry is to be omitted. These being closed in the
hull kernel topology, their complements, which are the same as
(\ref{5.14}), are open as asserted above.

The group $U$ is generated by transpositions $u(i,j)$ of
$h_i$ and $h_j$ for $ i \not= j$ :
\be
u(i, j) h_i = h_j~, ~~~u(i, j) h_j = h_i~, ~~~u(i, j) h_k =
h_k~~~{\rm if} ~~ k \not= i, j~.
\label{5.15}
\ee

Since the ideals of $n$ and $\infty$ are
\bea
& & I_n = \{1\!\!1 - \cp_n,\cp_1, \cp_2, \cdots, \Hat{\cp_n} , \cp_{n+1} ,
 \cdots \}~, \nonumber \\
& & I_{\infty} = \{\cp_1, \cp_2, \cdots \}~,
\label{5.16}
\eea
we find,
\bea
& & u(i, j)^{*} I_i u(i, j) = I_j~,~~~
u(i, j)^{*} I_j u(i, j) = I_i~, \nonumber \\
& &  u(i, j)^{*} I_k u(i, j) = I_k ~~~{\rm if}~~~k \not= i,j~,
\nonumber \\
& &  u(i, j)^{*} I_{\infty} u(i, j) = I_{\infty}~,
\label{5.17}
\eea
and
\bea
& & u(i, j) i = j~,~~~ u(i, j) j = i~,~~~ \nonumber \\
& & u(i, j) k = k~~{\rm if}~~~ k \not= i,j~, \nonumber \\
& & u(i, j) \infty = \infty~.
\label{5.18}
\eea

It is worth noting that the representation
(\ref{5.9}) of $\ca$ splits into a direct sum of the IRR's $p, q$
for the two-point poset. The proof
is as follows: $\infty$ being a fixed point for $U$, the functions
supported at
$\infty$ give an $\ca$-invariant one-dimensional subspace. It carries
the IRR $p$ by (\ref{4.1a}) and (\ref{5.13}). And since the orbit of
$n$ under $U$ is $\{1, 2, \cdots \}$, the functions vanishing at
$\infty$ give another invariant subspace. It carries the IRR $q$ by
(\ref{4.1a}) and (\ref{5.12}).

There is a suggestive interpretation of the projection operators
$\cp_n$. [See also the second paper of ref. \cite{Br}.]
The IRR $q$ of $\ca$ corresponds to the open set $[r, s[$
which restricted to $\cc$ splits into the direct sum of the IRR's~
$1, 2, \cdots $~. The IRR $p$ of $\ca$ corresponds to the point $s$
which restricted to $\cc$ remains IRR. We can think of $1, 2, \cdots $
{}~, as a subdivision of $[r, s[$ into points. Then $\cp_n$ can be
regarded as the restriction to $\Hat{\cc}$ of a smooth function on
$[r, s]$ with the value $1$ in a small neighbourhood of $n$ and the
value zero
at all $m \not= n$ and $\infty$.
In contrast, $1\!\!1$
is the function with value $1$ on the whole interval. Hence it has value
$1$ at all $n$ and $\infty$ as in (\ref{5.12}-\ref{5.13}). This
interpretation is illustrated in Fig 11.

\begin{figure}[hbtp]
\begin{center}
\large
\mbox{\psannotate{\psboxto(0cm;5cm){fig9.eps}}{ %\fillinggrid
\at(15\pscm;4\pscm){$q $}
\at(14\pscm;8\pscm){$p $}
\at(10.5\pscm;8\pscm){$\infty $}
\at(10.5\pscm;1\pscm){$1$}
\at(10\pscm;2.5\pscm){$2$}
\at(9.5\pscm;4\pscm){$3$}
}}
\normalsize
\end{center}
{\footnotesize{\bf Fig.11.}
The figure shows the division of $[r, s[$ into an infinity of
points $1, 2, \cdots $, which get increasingly dense towards $\infty$
or $p$. The point $\infty$, being a limit point of $1, 2, \cdots$, is
distinguished by a star. According to the suggested interpretation,
these points and $\infty$ correspond to IRR's of $\cc$ while $q =
\{1,2, \cdots \}$ and $p = \infty$ correspond to IRR's of $\ca$. }
\end{figure}

As mentioned previously, there is a certain difficulty in using the
scalar product (\ref{5.4}) for quantum physics.
For the two-point poset, it reads
\be
(g, h) = \sum_n g(n)^* h(n) + g(\infty)^* h(\infty)~,
\label{5.19}
\ee
where $\infty$ is the limiting point of $\{1, 2, \cdots \}$~.
Hence, if $h$ is a continuous function, and $h(\infty) \not=0$,
then Lim$_{n \rt \infty} h(n) = h(\infty) \not= 0$, and $(h, h) =
\infty$. It other words, continuous functions in $\ell^2(\Hat{\cc})$
must vanish at $\infty$. This is in particular true for probability
densities found from $\ce$. It is as though $\infty$ has been deleted
from the configuration space in so far as continuous wave functions
are concerned.

There are
two possible ways out of this difficulty.
a) We can
try regularization and modification of (\ref{5.4}) using some such
tool as the
Dixmier trace \cite{Co,VG}; b) we can try changing the scalar product
for example to $(\cdot~, ~\cdot)'_{\e}~, \e > 0$~, where
\be
(g, h)'_\e = \sum_n \frac{1}{n^{1 + \e}} g(n)^* h(n) + g(\infty)^*
h(\infty)~,
\label{5.20}
\ee
the choice of $\e$ being at our disposal.

There are minor changes in the choice of $u(i,j)$ if this scalar product
is adopted.

\subsxn{The $\bigvee$ Poset and General Two-Level Posets}\label{se:5.3}

In the case of the $\bigvee$ poset, there are Hilbert spaces $\ch_1$
and $\ch_2$ for each arm, $\ca$ being the algebra (\ref{4.2}) acting on
$\ch = \ch_1 \oplus \ch_2$. After choosing orthonormal basis
$h_n^{(i)}~, ~i = 1, 2~,~~n = 1, 2, \cdots $~, where the superscript
$i$ indicates that the basis element corresponds to $\ch_i$, and
orthogonal projectors $\cp_n^{(i)}$ on ${\bf C} h_n^{(i)}$, the
algebra $\cc$ can be written as
\be
\cc = < \cp_1, \bigcup_n \cp_n^{(1)} ;  \cp_2, \bigcup_n \cp_n^{(2)} >~.
\label{5.21}
\ee
Here $\cp_i$ are projection operators on $\ch_i$.

The group $U$ as before is generated by transpositions of basis
elements.

The space $\Hat{\cc}$ consists of two sequences $n^{(1)}~, n^{(2)}~
(n = 1, 2, \cdots $)~ and two points $\infty ^{(1)}$, $\infty ^{(2)}$,
with $n^{(i)}$ converging to $\infty^{(i)}$:
\be
\Hat{\cc} = \{ n^{(i)}~, \infty^{(i)}~; i = 1, 2 ~; n = 1, 2, \cdots
\}~. \label{5.22}
\ee
Their meaning is explained by
\bea
& & \cp_i (n^{(j)} ) = \delta_{ij}~, \nonumber \\
& & \cp_m^{(i)} (n^{(j)} ) = \delta_{ij} \delta_{mn}~,
\label{5.23}
\eea
\bea
& & \cp_i (\infty^{(j)} ) = \delta_{ij}~, \nonumber \\
& & \cp_m^{(i)} (\infty^{(j)} ) = 0 ~.
\label{5.24}
\eea

The visual representation of $\Hat{\cc}$ is presented in Fig. 12(a).

The remaining discussion of Section \ref{se:5.2} is readily carried out for
the $\bigvee$ poset as also for a general two-level poset.
So we content ourself by
showing the structure of $\Hat{\cc}$ for a $\bigvee\!\!\bigvee$ and a
circle poset in Figs. 12(b),(c).

\begin{figure}[hbtp]
\begin{center}
\large
\mbox{\psannotate{\psboxto(0cm;7.5cm){fig10a.eps}}{% \fillinggrid
\at(2\pscm;11.5\pscm){$\infty ^{(1)} $}
\at(8\pscm;11.5\pscm){$\infty ^{(2)} $}
\at(11\pscm;11.5\pscm){$\infty ^{(1)} $}
\at(17\pscm;11.5\pscm){$\infty ^{(2)} $}
\at(23\pscm;11.5\pscm){$\infty ^{(3)} $}
\at(4\pscm;1.5\pscm){(a)}
\at(16.5\pscm;1.5\pscm){(b)}
}}
\end{center}
\begin{center}
\mbox{\psannotate{\psboxto(0cm;7.5cm){fig10.eps}}{% \fillinggrid
\at(3\pscm;11.6\pscm){$\infty ^{(1)} $}
\at(11.5\pscm;11.6\pscm){$\infty ^{(2)} $}
\at(20.5\pscm;11.6\pscm){$\infty ^{(3)} $}
\at(12\pscm;0\pscm){(c)}
}}
\normalsize
\end{center}
\begin{center}
{\footnotesize{\bf Fig.12.}
The figures show the structure of  $\Hat{\cc}$ for three
typical two-level posets. }
\end{center}
\end{figure}

\subsxn{Abelianization from Gauge Invariance} \label{se:5.4}

The physical meaning of the algebra $U
\subset \ca$ is not very clear \cite{lisbon}, even though it is essential to
reproduce the poset as the structure space of $\ca$.

But if its role is just that and nothing more, is it possible to
reduce $\ca$ utilizing $U$ or a suitable subgroup of $U$ in some way
and get the algebra $\cc(\ca)$? The answer seems to be yes in all
interesting cases. We will now show this result and argue also that
this subgroup can be interpreted as a gauge group.

Let us start with the two-point poset. It is an ``uninteresting"
example for us where our method will not work, but it is a convenient
example to illustrate the ideas.

The condition we impose to reduce $\ca$ here is that the observables
must commute with $U$. The commutant $U'$ of $U$ in $\ca$ is just
${\bf C}1\!\!1$. The algebra $\ca$ thus gets reduced to a
commutative algebra, although it is not the algebra we want.

The next example is a ``good" one, it is the example of the $\bigvee$ poset.
The group $U$ here has two commuting subgroups $U^{(1)}$ and $U^{(2)}$.
$U^{(i)}$ is generated by the transpositions $u(k,l;i)$ which permute
only the basis elements $h_{k}^{(i)}$ and $h_{l}^{(i)}$:
\begin{eqnarray}
u(k,l;i) h_{k}^{(i)} & = & h_{l}^{(i)} ~,~u(k,l;i) h_{l}^{(i)} = h_{k}^{(i)}
{}~ ; \nonumber \\
u(k,l;i) h_{m}^{(j)} & = & h_{m}^{(j)} \mbox{ if } m \notin \{k,l\} ~
. \label{6.24}
\end{eqnarray}
These are thus operators acting along each arm of $\bigvee$, but do not act
across the arms of  $\bigvee$. The full group $U$ is generated by $U^{(1)}$
and $U^{(2)}$ and the elements transposing $h_{i}^{(1)}$ and $
h_{i}^{(2)}$.

Let us now require that the observables commute with $U^{(1)}$ and
$U^{(2)}$. They are given by the commutant of $\langle U^{(1)} ,
U^{(2)} \rangle$, the latter being
\be
\langle U^{(1)} , U^{(2)} \rangle ' = {{\bf C}} \cp_1 + {{\bf C}} \cp_2
\label{6.25}
\ee
in the notation of (\ref{4.2}). This algebra being isomorphic to
(\ref{6.23}), we get the result we want.

We can also find the correct representations to use in conjunction
with (\ref{6.25}). They are isomorphic to the IRR's of $\ca$ when
restricted to $\langle U^{(1)} , U^{(2)} \rangle '$. This is an obvious
result.

The procedure for finding the algebra $\cc(\ca)$ and its representations
of interest for a general poset now follows. Associated with each arm
$i$ of a poset, there is a subgroup $U^{(i)}$ of $U$. It permutes the
projections, or equivalently the IRR's [like the $n^{(i)}$ of
(\ref{5.22}) associated with this arm, while having the remaining
projectors, or the IRR's, as fixed points. The algebra $\cc(\ca)$ is
then the commutant of $\langle \bigcup_i U^{(i)} \rangle$ :
\be
\cc(\ca) = \langle \bigcup_i U^{(i)} \rangle ' ~ . \label{6.26}
\ee
The representations of $\cc(\ca)$ of interest are isomorphic to the
restrictions of IRR's of $\ca$ to $\cc(\ca)$.

In gauge theories, observables are required to commute with gauge
transformations. In an analogous manner, we here require the
observables to commute with the transformations generated by
$U^{(i)}$. The group generated by $U^{(i)}$ thus plays the role of
the gauge group in the approach outlined here.

\sxn{Final Remarks}\label{se:9}

In this article, we have described a physically well-motivated approximation
method to continuum physics based on partially ordered sets or posets. These
sets have the power to reproduce important topological features of continuum
physics with striking fidelity, and that too with just a few points.

In addition, there is also a remarkable
connection of posets to noncommutative geometry.
This connection comes about because
a poset can be thought of as a `noncommutative lattice',
being the dual space (the space of representations) of a
noncommutative algebra, and the latter is a basic algebraic
ingredient in noncommutative geometry.
The algebra of a poset also has a good intuitive meaning, being
the analogue of the algebra of
continuous functions on a topological space.

It is our impression that the above
connection is quite deep, and can lead to powerful and
novel schemes for numerical approximations which are also topologically
faithful. They seem in particular to be capable of describing solitons and
the analogues of QCD $\theta $-angles.

Much work of course remains to be
done, but there are already persuasive indications of the fruitfulness of the
ideas presented in this article for finite quantum physics.

\bigskip

\bigskip

\bigskip

\noindent
{\large \bf Acknowledgements }

This work was supported by the Department of Energy, U.S.A. under
contract number DE-FG02-ER40231. In addition, the work of G.L. was
partially supported by the Italian `Ministero dell' Universit\`a e
della Ricerca Scientifica'.

A.P.B. wishes to thank
Jos\'{e} Mour\~{a}o for very useful discussions and for drawing
attention to ref. \cite{Ma}. He thanks Cassio Sigaud for pointing out an error
in the preprint.

\end{document}